\documentclass[letterpaper,twocolumn,10pt]{article}
\usepackage{usenix-2020-09}

\usepackage{algorithm}
\usepackage{algpseudocode}

\usepackage{xspace}
\usepackage{subfigure}
\usepackage{url}
\usepackage{zref-totpages}

\usepackage{wrapfig}

\usepackage{hyperref}
\makeatletter
\def\UrlAlphabet{%
      \do\a\do\b\do\c\do\d\do\e\do\f\do\g\do\h\do\i\do\j%
      \do\k\do\l\do\m\do\n\do\o\do\p\do\q\do\r\do\s\do\t%
      \do\u\do\v\do\w\do\x\do\y\do\z\do\A\do\B\do\C\do\D%
      \do\E\do\F\do\G\do\H\do\I\do\J\do\K\do\L\do\M\do\N%
      \do\O\do\P\do\Q\do\R\do\S\do\T\do\U\do\V\do\W\do\X%
      \do\Y\do\Z}
\def\UrlDigits{\do\1\do\2\do\3\do\4\do\5\do\6\do\7\do\8\do\9\do\0}
\g@addto@macro{\UrlBreaks}{\UrlOrds}
\g@addto@macro{\UrlBreaks}{\UrlAlphabet}
\g@addto@macro{\UrlBreaks}{\UrlDigits}
\makeatother

\usepackage{tikz}

\usepackage{amsmath}
\usepackage{mathtools}

\newcommand{\name}{\textsc{Tarragon}\xspace}

\newcommand{\hide}[1] {}

\newcommand{\ie}{{\em i.e., \/}}
\newcommand{\eg}{{\em e.g., \/}}

\definecolor{lincolngreen}{rgb}{0.11, 0.35, 0.02}
\definecolor{majorelleblue}{rgb}{0.38, 0.31, 0.86}

\usepackage[normalem]{ulem}
\usepackage{comment}

\newcommand{\textnb}[1]{\noindent\textbf{#1}}

\usepackage{listings}

\definecolor{codegreen}{rgb}{0,0.6,0}
\definecolor{codegray}{rgb}{0.5,0.5,0.5}
\definecolor{codepurple}{rgb}{0.58,0,0.82}
\definecolor{backcolour}{rgb}{0.95,0.95,0.92}

\lstdefinestyle{mystyle}{
 backgroundcolor=\color{backcolour},
 commentstyle=\color{codegreen},
 keywordstyle=\color{magenta},
 numberstyle=\tiny\color{codegray},
 stringstyle=\color{codepurple},
 basicstyle=\ttfamily\footnotesize,
 breakatwhitespace=false,         
 breaklines=true,                 
 captionpos=b,                    
 keepspaces=true,                 
 numbers=left,                    
 numbersep=2pt,                  
 showspaces=false,                
 showstringspaces=false,
 showtabs=false,                  
 tabsize=2
}

\newcommand{\X}{$\times$\xspace}

\usepackage{multirow}
\usepackage{enumitem}
\setlist[enumerate]{leftmargin=0.5cm}
\setlist[itemize]{leftmargin=0.4cm}

\usepackage[T1]{fontenc}
\usepackage[breakable, theorems, skins]{tcolorbox}
\tcbset{enhanced}

\usepackage[textwidth=1cm]{todonotes}

\usepackage{booktabs}   
\usepackage{pifont}     
\usepackage{xcolor}     
\usepackage{graphicx}

\begin{document}

\date{}


\title{Making MoE-based LLM Inference Resilient with \name}

\author{
{\rm Songyu Zhang}\\
UC Riverside
\and
{\rm Aaron Tam}\\
UC Riverside
\and
{\rm Myungjin Lee}\\
Cisco Research
\and
{\rm Shixiong Qi}\\
University of Kentucky
\and
{\rm K. K. Ramakrishnan}\\
UC Riverside
} 


\maketitle

\begin{abstract}
Mixture‐of‐Experts (MoE) models are increasingly used to serve LLMs at scale, but failures become common as deployment scale grows.
Existing systems exhibit poor failure resilience:
even a single worker failure triggers a coarse-grained, service-wide restart, 
discarding accumulated progress and halting the entire inference pipeline %
during recovery---an approach clearly %
ill-suited for latency-sensitive, 
LLM services. 

We present \name, a resilient MoE inference framework that %
confines the failure's impact to individual workers 
while allowing the rest of the pipeline to continue making forward progress.
\name exploits the natural separation between the attention and expert computation in MoE-based transformers, treating attention workers (AWs) and expert workers (EWs) as distinct failure domains.
\name introduces a reconfigurable datapath to mask failures by rerouting requests to healthy workers.
On top of this datapath, \name implements a self-healing mechanism that relaxes the tightly synchronized %
execution of %
existing MoE frameworks.
For stateful AWs, \name performs asynchronous, incremental KV cache checkpointing with per-request restoration, and for stateless EWs, it leverages residual GPU memory to deploy shadow experts. These together keep recovery cost and recomputation overhead extremely low.
Our evaluation shows that, compared to state-of-the-art MegaScale-Infer, \name reduces failure-induced stalls by 160–213\X (from $\sim$64\,s down to 0.3–0.4\,s) while preserving performance when no failures occur.\footnote{We will open-source \name.}

\end{abstract}

\section{Introduction}
\vspace{-2mm}
\label{sec:intro}

Large Language Models (LLMs) have become a foundation of modern AI-powered applications (\eg conversation agents and code assistants).
However, as model sizes grow into hundreds of billions of parameters, serving them at low latency and at reasonable cost is increasingly challenging. 
The Mixture-of-Experts (MoE) architecture tackles this challenge by partitioning each dense feedforward network (FFN) layer
into many smaller ``experts'' and
using a ``gating network'' to activate only a subset of experts for each input token.
By avoiding full-model activation on every token, MoE enables LLMs to scale total parameter capacity and throughput without a proportional growth in per-token compute cost~\cite{outrageously-moe-2017, switch-transformers-jmlr22, megascle}.

As MoE serving scales to hundreds of GPUs, however, failures become increasingly common.
Measurements from large-scale GPU deployments show non-trivial failure rates: for example, \cite{cui2025characterizinggpuresilienceimpact} reports an average per-node uptime of 99.5\%, corresponding to around seven minutes of downtime per GPU node per day. So, in a 40-node (320 GPU) deployment like DeepSeek's~\cite{deepseekv3-tr}, this translates to roughly an 18.1\% probability (\ie $1-0.995^{40}$) of at least one node outage at any given time. Even worse, \cite{infiniteHBD} finds that
GPU-related faults alone account for 27.05\% of total system errors, representing the single largest failure source.

Unfortunately, today's MoE serving systems are extremely brittle under such failures.
In many systems, a single worker's failure triggers a \textit{coarse-grained} recovery: the entire inference job is torn down and restarted~\cite{llumnix, deepserve}, discarding all ``in-flight'' state (\eg partially constructed KV caches).
Here, a ``worker'' refers to a process running on a GPU in a node
participating in distributed inference, each typically executing a full transformer stack in monolithic serving %
engines (\eg vLLM~\cite{vllm}).
This strategy scales poorly, as restarting hundreds of workers
incurs long stalls, violates latency SLOs, and wastes computation already completed.
The root cause of this behavior lies in the tightly synchronized, monolithic execution model adopted by existing inference frameworks that rely on synchronous collective communications.
Thus, a single worker failure can trigger full restart---an approach ill-suited to interactive LLM services, where even sub-second disruptions degrade user experience.

\smallskip
\textnb{Alternative: Decoupled attention and expert computation.}
A key observation is that MoE inference inherently decomposes the forward pass into two distinct computational roles.
\textit{Attention} modules maintain a per-request key-value (KV) cache (initialized during the prefill phase) and append to it over time as decoding generates new tokens; 
In contrast, \textit{expert} modules implement FFN layers whose execution is stateless and depends only on incoming token embeddings and expert weights (this holds in both prefill and decoding phases).
This structural asymmetry suggests a natural way to define finer-grained failure domains:
attention-side computation is stateful and benefits from careful checkpointing, while expert-side computation is stateless and can be replayed or migrated.

Modern MoE serving systems increasingly decouple attention and expert computation across different workers,
rather than running a full transformer stack on each worker. This improves scalability and GPU utilization, %
as seen in systems such as MegaScale-Infer's Disaggregated Expert Parallelism (DEP)~\cite{megascale-infer}, DeepServe~\cite{deepserve}, and others %
with a similar decoupled architecture~\cite{hetermoe, step-3}.
We refer to this pattern as \emph{decoupled attention-expert deployment}, with Attention Workers (AWs) hosting attention modules and Expert Workers (EWs) hosting experts.
However, despite this structural decoupling, these systems still rely on \emph{tightly synchronized} execution between AWs and EWs. %
While this synchronous, batched execution %
achieves high GPU efficiency, it also \emph{globally exacerbates the impact of failures} by 
stalling the entire inference pipeline.

\smallskip
\textnb{Our Approach: \name.} 
Building on these observations, we present \name, a resilient MoE inference framework that achieves stall-free, fine-grained failure recovery with minimal performance overhead.
Rather than treating the entire inference job as a single failure domain, \name fully exploits the decoupled attention-expert deployment paradigm and separates AWs and EWs into distinct failure domains.
The core objective of \name is: 
\textit{When a failure occurs, \name confines its impact to the corresponding domain instead of restarting the entire job, allowing the rest of the inference pipeline to keep making forward progress without disruption.}

At a high level, \name makes two complementary design choices to serve its core objective.
First, \name realizes distinct failure domains through a reconfigurable datapath between AWs and EWs.
This datapath is implemented by
a Reconfigurable Forwarding Engine (REFE) %
that mediates all AW-EW communication with an Expert Routing Table (ERT) that
dynamically binds each \emph{expert identity} (\ie the logical expert selected by the gating network) to \emph{expert location} (\ie the physical EW/GPU hosting that expert).
In many existing systems,
this binding is static---each logical expert is permanently tied to a specific EW. As a result, when that EW fails, the exper itself becomes unavailable, forcing a pipeline-wide restart.
In contrast, \name eliminates fixed bindings: AWs issue requests only in terms of logical expert IDs; REFE consults the ERT to resolve where each expert currently resides and routes the requests accordingly.

Second, to achieve non-disruptive failover, \name introduces a 
self-healing mechanism combined with new worker provisioning in parallel.
Both mechanisms critically rely on having a failure domain at the granularity of an individual worker.
Upon an AW or EW failure, self-healing keeps the pipeline running by reacting locally within each domain: quickly moving affected requests off failed workers onto healthy ones, so that the inference pipeline does not pause waiting for global recovery.
In parallel, the \name's control plane provisions replacement AWs/EWs and integrates them into the ongoing inference pipeline, restoring lost capacity.

Furthermore, to minimize the cost of recovery, \name tailors its failure resilience strategy to the different roles of AWs and EWs across the prefill and decoding phases.
Because EWs are stateless in both phases (as discussed above), EW failures can be handled purely via replay on healthy EWs, %
accelerated by ``shadow experts'' that occupy available %
GPU memory but consume no compute resources. %
This design simplifies recovery to just reusing GPU-resident expert weights, avoiding costly reloads from storage. %

For AWs, however, the cost of failure is phase-dependent.
During prefill, recovery requires recomputing the KV cache from the prompt, which incurs extra work but does not disrupt an ongoing interaction. 
During decoding, by contrast, rebuilding the KV cache by replaying the full token history would introduce prohibitive latency.
As we show in \S\ref{sec:anatomy-coarse-failures}, prefill failures are relatively cheap, while decoding-time failures dominate recovery cost and are therefore our main optimization focus.
To avoid this decoding phase penalty,
\name performs \emph{asynchronous, incremental} KV cache checkpointing and per-request restoration.
\name also exploits short idle gaps on the AW-EW datapath to avoid interfering with normal traffic.
This enables \name to resume decoding from the latest emitted token after an AW failure while keeping checkpoint bandwidth and compute overhead modest.

\smallskip
\textnb{Contributions.}
This paper makes the following contributions:

\begin{itemize} [leftmargin=10pt,nosep]
    \item We analyze why existing MoE inference systems are brittle under failures (\S\ref{sec:anatomy-coarse-failures}), showing how synchronous dependencies between AWs and EWs cause a single worker failure to stall the entire inference pipeline.

    \item End-to-end, \name's self-healing combined with new worker provisioning reduces failure-induced stalls from $\sim$64\,s in a MegaScale-like baseline to 0.4\,s for AW failures and 0.3\,s for EW failures (160–213$\times$ improvement; \S\ref{sec:e2e-failover}).

    \item Under no failures, \name closely matches MegaScale-Infer in throughput and token-level latency (within 2.8\%; \S\ref{sec:cost-of-failure-resiliency}); under failures, it achieves the above stall reductions without sacrificing steady-state performance.

    \item The asynchronous, incremental KV cache checkpointing has negligible overhead ($<3\%$) in the AW-EW datapath. The per-request restoration is able to reduce AW recovery latency by up to 1800\X, recovery traffic by up to 8\X, and eliminates any GPU recomputation that would be needed if we were to rebuild the entire KV cache.

\end{itemize}

\section{Background and Motivation}
\vspace{-2mm}
\label{sec:background-motivation}

\subsection{Basics of MoE Inference}
\vspace{-2mm}
\label{sec:moe-basics}
Transformers form the foundation of modern LLMs, comprising multiple self-attention and feed-forward network (FFN) layers stacked  to improve compositional generalization and language modeling performance~\cite{petty-2024-impact}.
In conventional (dense) transformers, every token is processed by all FFNs in each transformer block. 
This dense activation causes both computation and memory costs to scale linearly with model size, making large models expensive to serve~\cite{switch-transformers-jmlr22}.

\begin{figure}[t]
\centering
    \includegraphics[width=\linewidth]{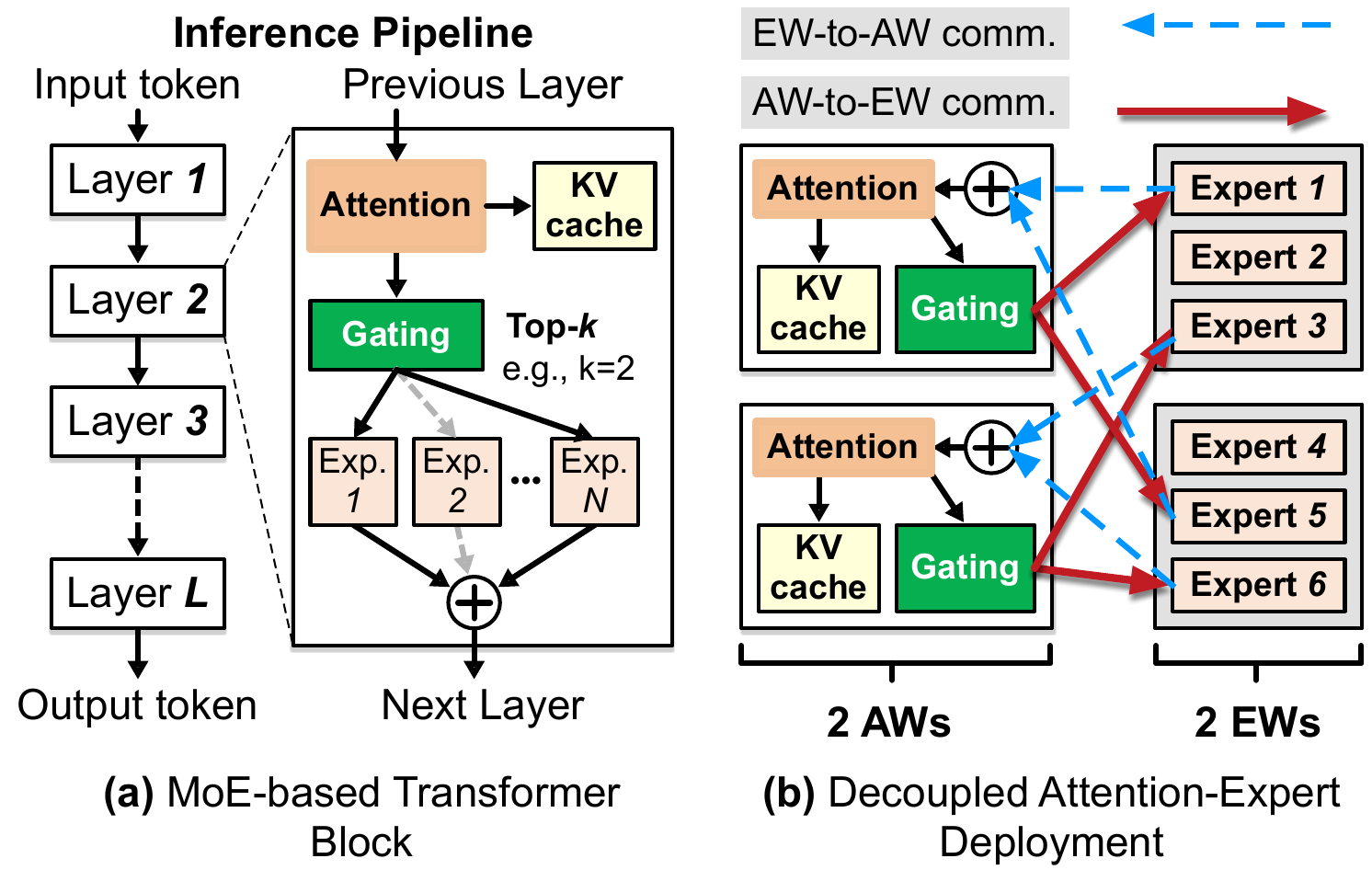}
\vspace{-8mm}
    \caption{(a) MoE-based transformer layer and LLM inference pipeline: example shows top-$2$ experts selected; (b) decoupled attention-expert deployment.}
    \label{fig:moe-overview}
\vspace{-3mm}
\end{figure}

To address this limitation, recent LLMs adopt MoE-based models, which
replace the dense FFN layer with a sparse MoE layer~\cite{deepseekai2025deepseekv32pushingfrontieropen, qwen3technicalreport,llama4,openai2025gptoss120bgptoss20bmodel,kimiteam2025kimik2openagentic}.
As shown in Fig.~\ref{fig:moe-overview}(a), an MoE layer consists of multiple FFNs (called \textit{experts}) and a \textit{gating network} that selects only the top-$k$ experts for each token, typically with a small expert-selection ratio.%
\footnote{For instance, only 8 of 256 experts are activated in DeepSeek-v3~\cite{deepseekv3-tr}, and 8 of 128 in Qwen3-MoE~\cite{qwen3technicalreport}.}
Each selected expert processes the token independently, and the resulting expert outputs are aggregated via a weighted sum using the gating weights before being passed to the next layer.

By activating a small subset of experts per token, MoE models increase total parameter capacity without proportionally increasing per-token compute cost, making them attractive for \emph{inference} workloads~\cite{lina,megascale-infer,deepseek-moe}.
In this work, we focus exclusively on MoE-based LLM \emph{inference}; training-specific concerns
are beyond the scope of this paper.

\textnb{Prefill versus decoding.}
Inference naturally decomposes into two phases.
During the \emph{prefill} phase, the model consumes an input prompt (often hundreds or thousands of tokens) and builds the internal context state (KV cache).
Prefill leverages substantial parallelism because tokens in the input sequence are all independent inputs~\cite{Sarathi-serve}.
During the \emph{decoding} phase, the model generates output tokens sequentially, one token at a time. Each new token depends on all previously generated tokens, forcing decoding to proceed in a strictly sequential manner at single-token granularity~\cite{Sarathi-serve}. %
Despite their distinct execution characteristics, both phases traverse the same stack of transformer layers.

\textnb{Stateful attention versus stateless expert.}
Transformers exhibit a fundamental asymmetry in how state is managed across attention and expert components.
On the attention side, the model maintains a per-request KV cache that stores the key and value projections for all previously processed tokens.
This cache is initialized during prefill and incrementally extended during decoding as new tokens are generated.
As a result, attention computation is \emph{stateful}: %
its execution depends on a growing context that persists across layers and across time.

In contrast, experts are stateless FFNs with fixed weights and no per-request persistent state. %
Given a batch of token embeddings, an expert's output depends only on its input activations and its static parameters.
This stateless property holds in both prefill and decoding: expert computation is a pure function that can always be reproduced by replaying the same inputs.

\begin{figure}[b]
\vspace{-6mm}
\centering
    \includegraphics[width=\linewidth]{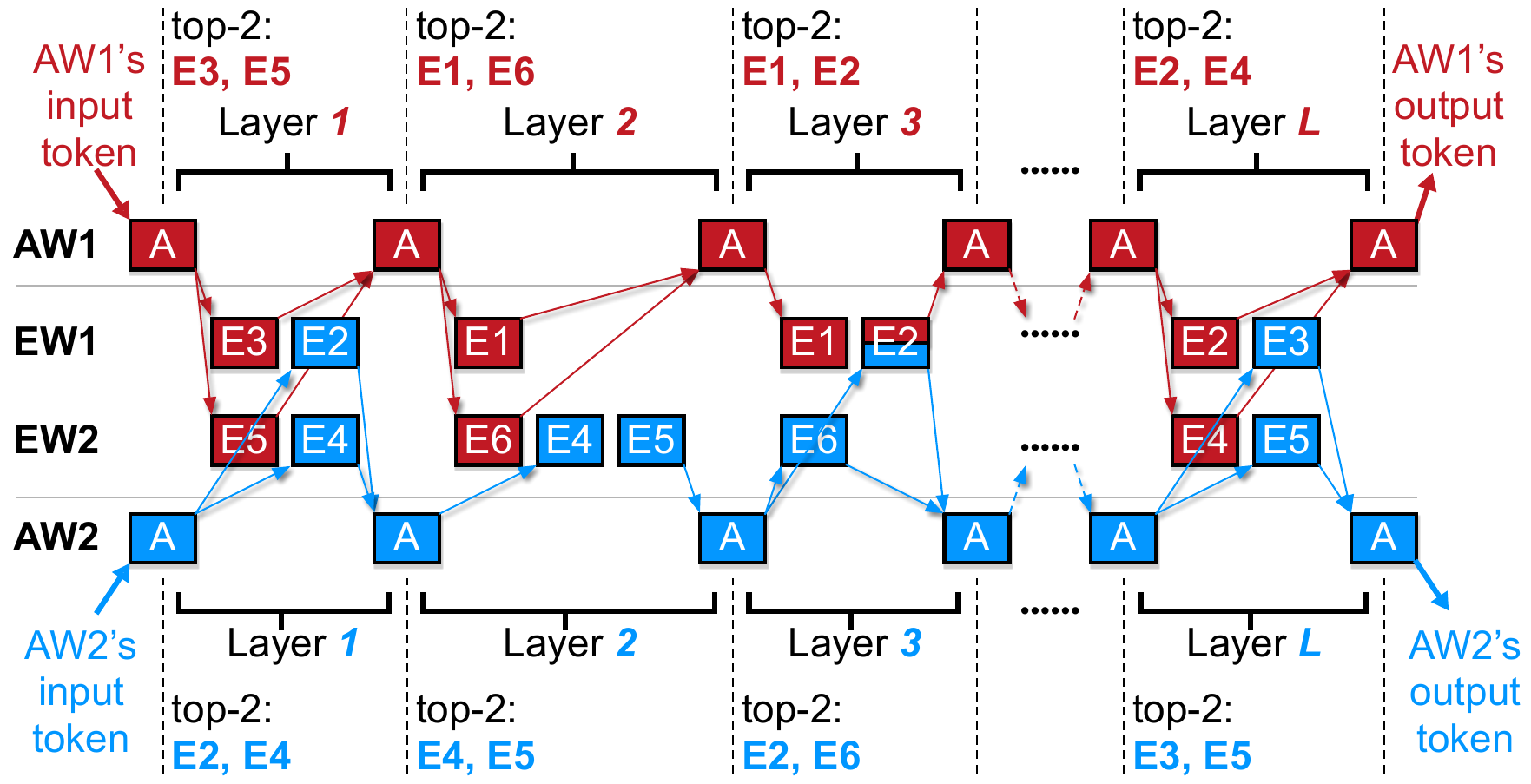}
\vspace{-8mm}
    \caption{Example of layer-wise synchronized MoE inference. Here we show two data-parallel AWs, and two EWs, each hosting three expert FFNs (E1-E6), the same as Fig.~\ref{fig:moe-overview}(b).}

    \label{fig:sync-barrier}
\end{figure}

\textnb{Deployment patterns for MoE inference.}
Early MoE serving systems often adopted a \emph{monolithic} deployment model, where a single worker process (typically bound to one or more GPUs) hosts an entire transformer stack, including both attention and expert modules~\cite{vllm,sglang}.
Workers communicate %
via collectives (\eg NCCL's all-to-all~\cite{nccl}).
While simple to implement, this design scales poorly in terms of memory efficiency and GPU utilization~\cite{megascale-infer}.

To address these issues, recent production systems~\cite{megascale-infer, step-3}
have shifted to %
\emph{decoupled attention-expert deployment}, in which attention and expert modules are placed on separate sets of workers (Fig.~\ref{fig:moe-overview}(b)).
We refer to these as Attention Workers (AWs) and Expert Workers (EWs).
This separation enables independent scaling of AWs and EWs, allowing expert traffic from many AWs to be consolidated onto fewer EWs 
to improve batching efficiency and GPU utilization.

\begin{figure*}[htbp]
\centering
    \includegraphics[width=\linewidth]{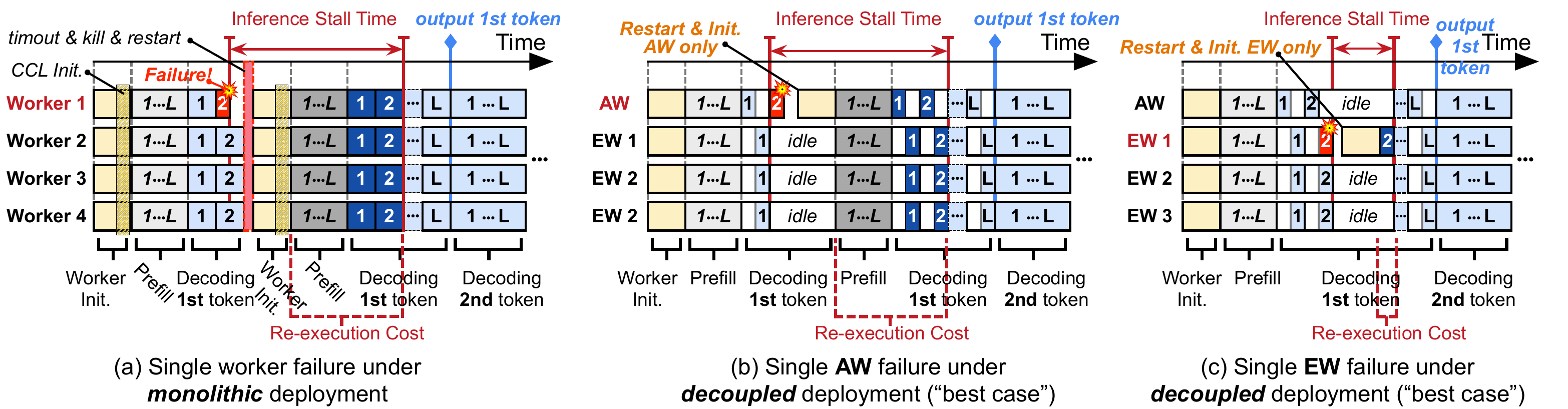}
\vspace{-8mm}
    \caption{Coarse‐grained failure recovery under different deployment modes.
    Numbers inside the ``decoding'' boxes denote 
    the transformer layer currently being executed. 
    For the decoupled deployment, we analyze a ``best-case'' recovery scenario, where a single worker failure (AW or EW) results only in the failed worker restarting. However, some existing decoupled systems still restart all workers on failure, thus effectively degenerating to the monolithic case.
    }
    \label{fig:failure-case-studies}
\vspace{-5mm}
\end{figure*}
In practice, AWs are typically scaled out using data parallelism~\cite{megascle,step-3}, with each AW serving a disjoint subset of requests, while EWs form an
expert-parallel group that partitions expert FFNs across GPUs.
Because AW-EW traffic follows an asymmetric many-to-many pattern rather than a fixed symmetric collective, standard CCLs such as NCCL's all-to-all are a poor fit~\cite{megascle,step-3}. Recent systems therefore employ custom AW-EW data planes (\eg MegaScale-Infer's M2N) to support flexible expert routing and elastic scaling~\cite{megascle,step-3}.

\subsection{Anatomy of Coarse-Grained Failures}
\vspace{-2mm}
\label{sec:anatomy-coarse-failures}

To motivate 
the need for fine-grained, failure-resilient MoE inference,
we first explain how layer-wise synchronization governs MoE execution, then analyze failure propagation under two representative deployment modes, and finally quantify the resulting recovery overheads.

\subsubsection{Layer-wise synchronized execution}
\vspace{-2mm}
\label{sec:layer-wise-execution}
As described in \S\ref{sec:moe-basics}, MoE inference advances layer by layer under a strict synchronization barrier between attention and expert computation.
In decoupled deployments, this execution is distributed across AWs and EWs, as illustrated in Fig.~\ref{fig:sync-barrier}.

Two important properties follow: 
(1) For each layer $\ell$, every data-parallel AW independently processes its own request, 
selects a 
subset of experts, sends token embeddings to the corresponding EWs, and \textit{waits until all selected experts return their outputs} (a synchronization barrier) before advancing to layer $\ell+1$.
We refer to the current layer index $\ell$ of an AW as its \textit{frontier};
(2) On the EW side, GPUs execute expert FFNs in \emph{layer-wise batches}: an EW aggregates requests for the same layer $\ell$ and expert, and executes them as a single large batch, and only then advances to the next layer. This layer-wise batching effectively ties EW progress to the same frontier as the AWs and is crucial for GPU efficiency~\cite{megascale-infer}; naively executing each request immediately upon arrival destroys batching opportunities and severely underutilizes GPUs~\cite{megascale-infer}.

This layer-wise synchronization pattern holds across both prefill and decoding, and across both monolithic and decoupled deployments. 
Further, this layer-wise barrier is not tied to a particular parallelization scheme.
Even under tensor or pipeline parallelism, each data-parallel AW group behaves as a logical ``mega-AW'' 
with the same layer-wise barrier.

\subsubsection{Case studies and quantifying recovery overheads}
\vspace{-2mm}
\label{sec:cost-analysis}

We now examine how a single worker's failure propagates under two representative MoE deployment modes discussed above.

We consider a failure during \emph{decoding} an $L$-layer MoE model, while
generating the $i$-th output token and executing layer $\ell$ ($1 \le \ell \le L$).
This setting captures the worst %
impact on user-perceived latency, since the request has
already gone through prefill and is in the middle of token generation.

Fig.~\ref{fig:failure-case-studies}(a)–(c) shows how a single worker failure escalates into a coarse-grained disruption.
\footnote{We provide a detailed analysis of Fig.~\ref{fig:failure-case-studies}(a)–(c) in Appendix~\ref{sec:case-studies-walk-through}.}
In all three cases, a single worker failure effectively induces a service-wide stall and the recovery incurs two fundamental penalties (highlighted in Fig.~\ref{fig:failure-case-studies}): 
(1) \emph{Inference stall time} ($T_{\text{stall}}$), the duration during which the pipeline cannot generate new tokens for the affected request;
(2) \emph{Re-execution cost} ($G$), the amount of wasted GPU time/cycles required to recompute lost work.
We now build a cost model
to understand, \emph{for a fixed model and deployment configuration}, how the \emph{failure point} (captured by the decoded-token index $i$ and the frontier layer~$\ell$) affects recovery cost.
For clarity, we assume that workers are perfectly load-balanced and have comparable per-layer performance (thus, we ignore stragglers to keep the model simple).

\textnb{Inference stall time.}
Let $T_w$ be the average time to (re)initialize a worker, %
including process (or container) startup, CUDA context initialization, %
loading weights, and %
communication stack initialization~\cite{serverlessLLM, hydraserve-nsdi26}. 
Let $t_{\text{pre}}$ and $t_{\text{dec}}$ be the average execution time %
of one prefill layer and  one decoding layer for a single token, respectively. %
As shown in Fig.~\ref{fig:failure-case-studies}(a) and (b), 
the recovery procedure for a monolithic worker and for a decoupled AW has the same structure:
the failed worker is restarted, then all workers must replay all prefill and decoding layers up to the failure point $(i,\ell)$. 
In the monolithic deployment, a single worker failure also kills all healthy workers, as the collective communicator (CCL) treats the worker set as a static communication group and aborts when any worker is lost~\cite{Hoplite}.%
\footnote{In practice, this often manifests as a fatal NCCL/MPI error that terminates the job~\cite{Hoplite}.}
Ignoring lower-order effects (\eg warm caches, overlap), the stall time can thus be approximated as:

\vspace{-6mm}
\begin{equation}
  T_{\text{stall}}(\ell, i)
  ~\approx~
  \underbracket[0.8pt][3pt]{T_w}_{\raisebox{-0.8em}{$\substack{\text{Worker}\\\text{Reinit.}}$}}
  ~+~
  \underbracket[0.8pt][3pt]{L \cdot t_{\text{pre}}}_{\raisebox{-0.8em}{$\substack{\text{Replay $L$}\\\text{prefill layers}}$}}
  ~+~
  \underbracket[0.8pt][3pt]{\bigl[(i-1) L + \ell\bigr] \cdot t_{\text{dec}}}_{\raisebox{-0.8em}{$\substack{\text{Replay decoding up to}\\\text{layer $\ell$ of the $i$-th token}}$}}
  \label{eq:stall-mono-aw}
\end{equation}
\vspace{-3mm}

\noindent
For a decoupled EW failure (Fig.~\ref{fig:failure-case-studies}(c)), 
prior prefill and decoding work is preserved on the AWs since EWs are stateless.
Recovery only requires reinitializing the EW and re-executing the expert layer at the current frontier:

\vspace{-3mm}
\begin{equation}
  T_{\text{stall}}(\ell, i)
  ~\approx~
  \underbracket[0.8pt][3pt]{T_w}_{\raisebox{-0.8em}{$\substack{\text{Worker}\\\text{Reinit.}}$}}
  ~+~
  \underbracket[0.8pt][3pt]{t_{\text{dec}}}_{\raisebox{-0.8em}{$\substack{\text{Replay decoding}\\\text{at frontier $\ell$}}$}}
  \label{eq:stall-ew}
\end{equation}
\vspace{-3mm}

\textnb{Re-execution cost (GPU computation overhead).}
We measure GPU overhead in units of \emph{GPU-time}, defined as the product of execution time and the number of GPUs (SMs) performing that recomputation.
Let $g_{\text{pre}}$ and $g_{\text{dec}}$ denote the average per-worker GPU-time cost of processing one prefill layer and one decoding layer for a single token, respectively.
For a monolithic deployment with $M$ workers, all workers must replay the lost computation.
Again, the decoupled AW failure follows the same replay pattern.
The total GPU computation overhead is therefore:

\vspace{-6mm}
\begin{equation}
  G(\ell, i)
  ~\approx~
  M \cdot \bigl[
    \underbracket[0.8pt][3pt]{L \cdot g_{\text{pre}}}_{\raisebox{-0.8em}{$\substack{\text{Replay $L$}\\\text{prefill layers}}$}}
    ~+~
    \underbracket[0.8pt][3pt]{\bigl((i-1) L + \ell\bigr) \cdot g_{\text{dec}}}_{\raisebox{-0.8em}{$\substack{\text{Replay decoding up to}\\\text{layer $\ell$ of the $i$-th token}}$}}
  \bigr]
  \label{eq:gpu-waste-mono-aw}
\end{equation}
\vspace{-4mm}

\noindent
For a failed EW, only the expert computation at the current frontier must be repeated on a single replacement EW: %

\vspace{-3mm}
\begin{equation}
  G(\ell, i)
  ~\approx~
  g_{\text{dec}}
  \label{eq:gpu-waste-ew}
\end{equation}
\vspace{-6mm}

\smallskip
\textnb{Experimental Setup for Audit.}
We validate this cost model using measurements from (1) a monolithic vLLM deployment and (2) a decoupled MegaScale-Infer-like deployment (configurations detailed in \S\ref{sec:expt-setup}).
Both systems serve the Mixtral-8$\times$7B model (32 layers).
The number of workers is 16 (8~AWs \& 8~EWs for decoupled deployment).
For each configuration, we empirically measure $T_w$, $t_{\text{pre}}$, $t_{\text{dec}}$,
$g_{\text{pre}}$, and $g_{\text{dec}}$ (see Table~\ref{tab:empirical-params}).
We then sweep the failed decoded-token index $i$
to evaluate recovery overhead across different decoding stages.

\begin{table}[b]
\vspace{-6mm}
  \centering
  \caption{Profiled parameters for the overhead analysis.}
  \label{tab:empirical-params}
  \resizebox{\linewidth}{!}{%
  \begin{tabular}{lccccc}
    \toprule
    \textbf{Deployment} 
      & $T_w$ 
      & $t_{\text{pre}}$ 
      & $t_{\text{dec}}$ 
      & $g_{\text{pre}}$ 
      & $g_{\text{dec}}$ \\
    \midrule
    vLLM~\cite{vllm}     
      & 24~s & 1.68~ms & 0.58~ms & 0.010 & 0.0028 \\
    MegaScale-Infer~\cite{megascale-infer} 
      & 18.5~s & 2.18~ms & 0.85~ms & 0.006 & 0.0022 \\
    \bottomrule
  \end{tabular}
  }
\end{table}

\begin{figure}[t]
\centering
    \includegraphics[width=\linewidth]{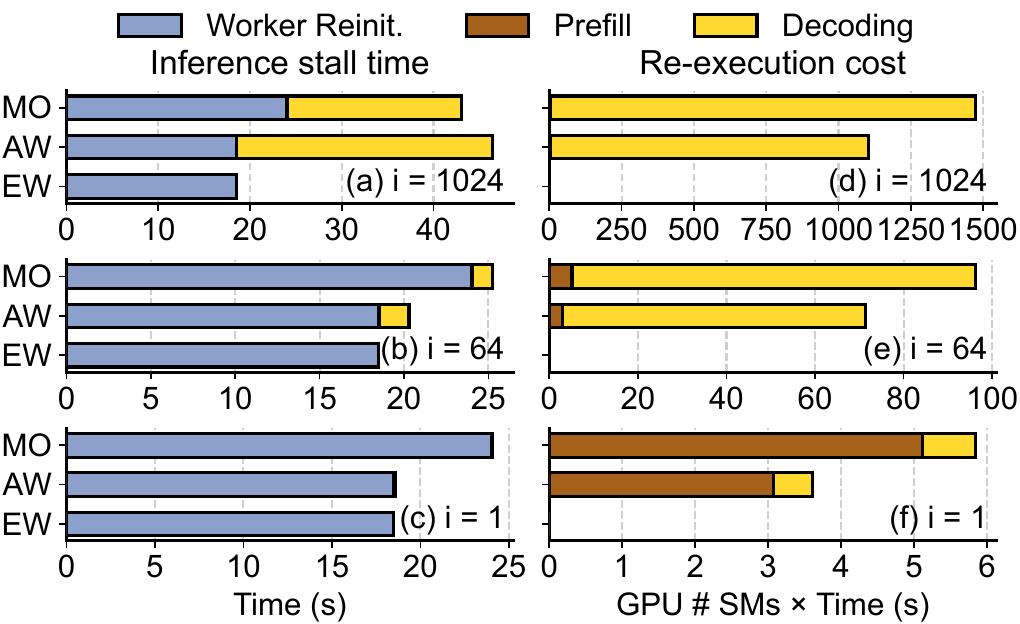}
\vspace{-8mm}
    \caption{(a–c) Inference stall time and (d–f) re-execution cost 
    under a single worker failure. MO: monolithic worker.
    }
    \label{res:failure_cost_heatmaps}
\vspace{-5mm}
\end{figure}

Fig.~\ref{res:failure_cost_heatmaps} shows the inference stall time and the re-execution cost,
which reveals three observations: 
\textbf{(1)} For the failure of the monolithic worker and decoupled AW, stall time and wasted computation grow rapidly with the decoded-token index $i$:
a failure later during decoding forces replay of a long history
even when only one worker fails; 
\textbf{(2)} Decoding-time failures are the dominant concern: even when only 64 tokens have been decoded, the recovery cost in decoding already exceeds that of a prefill failure with a 128-token prompt by about 19\X, 
highlighting decoding as the primary target for optimization;
\textbf{(3)} Although decoupled deployments limit the EW failure to single-layer re-execution, which introduces only a constant-time stall and the GPU overhead,
the worker initialization cost $T_w$ still remains on the critical path. Thus, EW failures can still introduce user-visible pauses even though the replay expense is small.

\subsubsection{Takeaways}
\vspace{-2mm}
\label{sec:motivation-takeaways}

This analysis reveals three fundamental problems with the current failure handling in MoE inference:

\begin{itemize}[leftmargin=10pt,nosep]
  \item \textit{Overly coarse failure domains.}
  Regardless of deployment modes, a single worker failure effectively enlarges the failure domain to the level of the full inference service, forcing all participating workers to restart or at least wait.

  \item \textit{User-visible stalls.}
  Because the worker(s) must be restarted ($T_w$) and redo prefill and decoding before emitting new output tokens, recovery delays propagate directly to interactive users as broken conversational flows.

  \item \textit{Wasted computation.}
  Previously computed KV cache and expert outputs are discarded and recomputed from scratch; the longer a request has been decoding when a failure occurs, the more GPU time is wasted.
\end{itemize}
Our goal in the rest of this paper is to design a MoE inference system that has (\textbf{D1}) \emph{fine-grained worker-granularity failure domains}, (\textbf{D2}) \emph{minimizes failure-induced stalls}, and (\textbf{D3}) \emph{preserves as much useful computation as possible}.

\section{Overview of \name}
\vspace{-2mm}
\label{sec:design}

\subsection{High-level Approach}
\vspace{-2mm}
\label{sec:high-level-ideas}

\name achieves three major goals %
by rethinking how failures interact with the
decoupled attention-expert %
deployment. %

\smallskip
\textnb{(D1): Reconfigurable worker-level failure domains.}
Although decoupled attention-expert deployments naturally suggest worker-granularity failure domains, existing systems fail to realize this because placement and AW-EW routing are statically bound: each expert is pinned to a fixed GPU, and routing is baked into the datapath.
\name breaks this static coupling and realizes \textbf{D1} %
with a \emph{reconfigurable} AW-EW datapath (\S\ref{sec:datapath}).
Each AW dispatches requests through a Reconfigurable Forwarding Engine (REFE) that resolves logical expert IDs to physical EWs via
an Expert Routing Table (ERT) (\S\ref{sec:refe-design}).
Upon failure, the orchestrator updates the ERT to redirect traffic to healthy EWs without restarting AWs or pausing the pipeline. 
Meanwhile, EWs accept traffic from any AW without joining or recreating a collective group.

\smallskip
\textnb{(D2): Self-healing and background capacity restoration.}
While reconfigurable routing prevents global restarts, it does not eliminate stalls ($T_{\text{stall}}(\ell,i)$ quantified in \S\ref{sec:cost-analysis}) caused by worker reinitialization and replay.
To minimize such stalls, \name layers self-healing and background capacity restoration %
on top of its datapath.
\emph{Self-healing} decouples pipeline progress from worker recovery:
when a worker (AW or EW) fails, its requests are immediately replayed on healthy alternatives rather than blocking until the failed worker reboots 
(\S\ref{sec:aw-side-self-healing}, \S\ref{sec:ew-side-self-healing}). 
For EWs, failover is further accelerated through shadow experts (\S\ref{sec:shadow-experts-design}), which are pre-loaded into residual GPU memory, so rerouting avoids costly reloads. 
In parallel, the orchestrator performs \emph{background worker reprovisioning} (\S\ref{sec:new-worker-provision}) to restore lost capacity without interrupting active inference.
Together, these two mechanisms tackle the two dominant contributors to stall time identified in \S\ref{sec:cost-analysis}---worker restart delay $T_w$ and long replay paths.

\smallskip
\textnb{(D3): Efficient recovery for stateful AWs.}
Our cost analysis (\S\ref{sec:cost-analysis}) shows that \emph{late} AW failures are especially expensive because KV caches must otherwise be rebuilt through replay (see Eq.~\eqref{eq:stall-mono-aw} and~\eqref{eq:gpu-waste-mono-aw}), adding stall time and GPU overhead. %
During this recovery, the attention module appends a small, fixed-size KV cache (denoted ``segment'' hereafter) for each token.
\name mitigates this by introducing \emph{asynchronous, incremental KV cache checkpointing} (\S\ref{sec:async_checkpoint}) and \emph{request-level KV cache restoration} (\S\ref{sec:kv-restoration}).
AWs continuously stream newly appended KV cache 
segments
to an external checkpoint store without interfering with the inference pipeline.
Upon failure, the orchestrator only restores the affected requests' KV caches on healthy AWs, shrinking recovery to ``roughly one decoding layer at the current frontier'' instead of replaying full prefill and decoding. This dramatically reduces both stall time ($T_{\text{stall}}$) and GPU recomputation ($G$).

\begin{figure}[t]
\centering
    \includegraphics[width=\columnwidth]{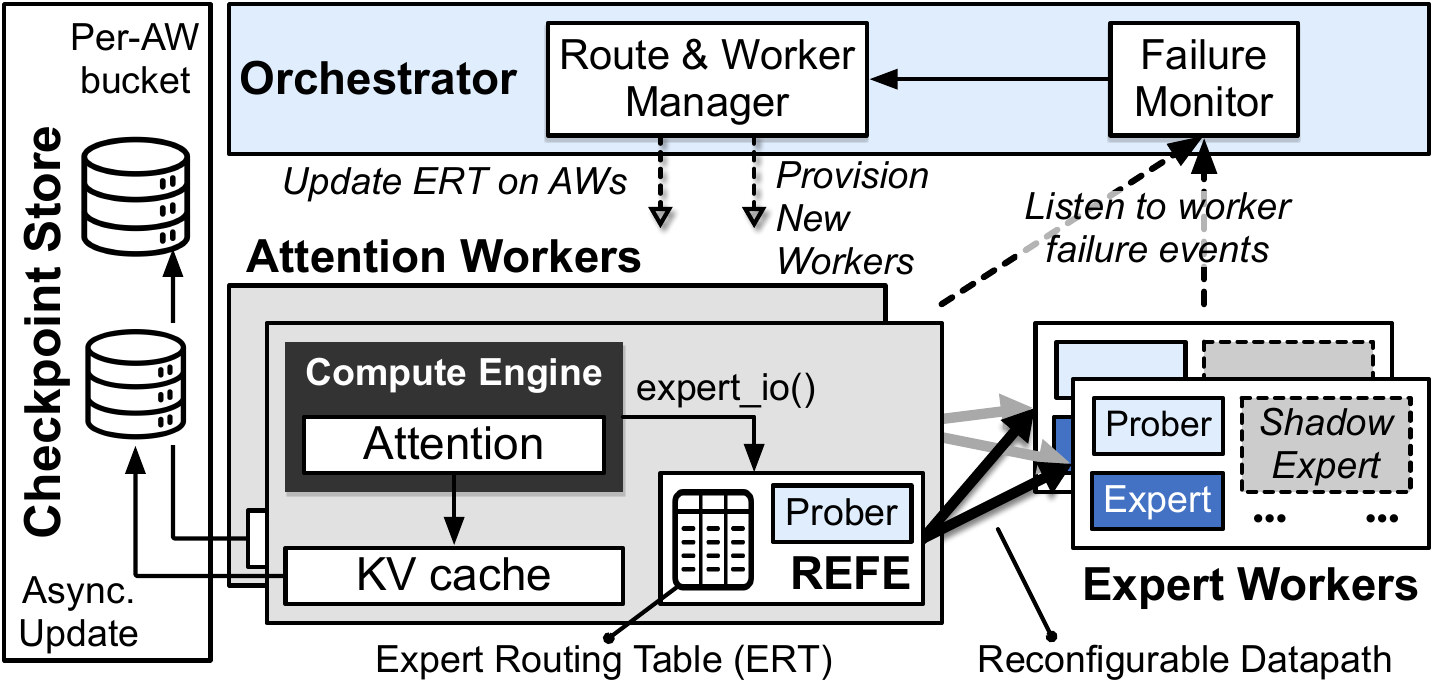}
\vspace{-8mm}
    \caption{Overview of \name. 
    }
\vspace{-6mm}
    \label{fig:overview}
\end{figure}

\subsection{\name Architecture}
\vspace{-2mm}
\label{sec:overview}

Fig.~\ref{fig:overview} summarizes \name's architecture.
Every AW has a Compute Engine (built atop vLLM) that hosts attention modules, manages the per-request KV cache, and interacts with EWs through the REFE, which implements the reconfigurable AW-EW datapath using the ERT.
A single cluster gateway distributes user requests to the AWs.
EWs host active and shadow experts.
A centralized orchestrator monitors worker liveness for failure detection, includes a manager that updates the ERT on failures and joins, and coordinates background provisioning of new AWs and EWs.
A checkpoint store receives incremental KV cache updates
from AWs and serves request-level state to replacement AWs during recovery.

\subsection{Failure Model}
\vspace{-2mm}
\label{sec:failure-model}

\name adopts a fail-stop failure model focused on hardware and software crashes. We assume workers (AWs and EWs) may fail due to application or OS crashes, node failures, power outages, and other unplanned interruptions. Of particular concern are CPU and GPU failures or errors requiring either a restart or a repair of the node. In practice, MoE serving clusters are GPU-heavy, so GPU device error is the dominant class of failures we target~\cite{cui2025characterizinggpuresilienceimpact, infiniteHBD}.
\name also treats communication link failures as fail-stop events. 
Modern GPU clusters frequently experience link-level faults, \eg fabric-level connectivity loss and intra-node PCIe/NVLink disruption~\cite{infiniteHBD,Fire-fly-hpc,Characterization-LLM}, 
which effectively isolate a worker even if its process is still running. In such cases, \name considers the affected worker unreachable and handles it similarly to a fail-stop crash.
Byzantine failures are explicitly out of scope for this paper. Our detection and recovery mechanisms assume components are not malicious and that all the nodes (AW, EW, orchestrator) have a consistent (even if delayed) view of component failures. %

\section{Datapath for Fine-grained Failure Domains}
\vspace{-2mm}
\label{sec:datapath}

\begin{figure}[b]
\vspace{-4mm}
\centering
    \includegraphics[width=\columnwidth]{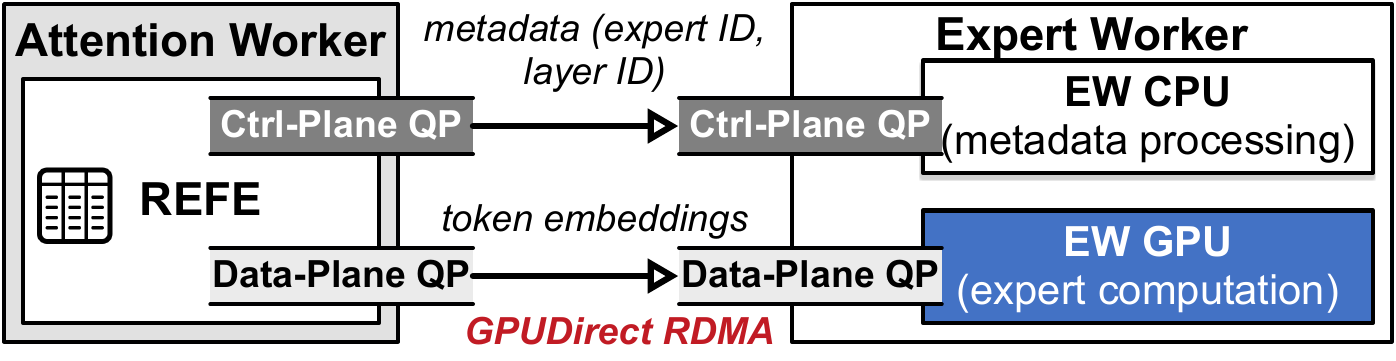}
\vspace{-8mm}
    \caption{AW-EW datapath in \name. AWs dispatch requests to EWs through the REFE, which separates metadata and tensor transfers across two RDMA QPs.}
    \label{fig:aw-ew-datapath}
\end{figure}

The reconfigurable datapath in \name isolates failures at the worker level.
As shown in Fig.~\ref{fig:aw-ew-datapath}, its key design aspects include the separation of control metadata from high-volume tensor transfers (\S\ref{sec:split-cp-dp}),
and the REFE that dynamically reroutes requests to healthy EWs upon failure detection (\S\ref{sec:refe-design}). 
\name currently implements its datapath using RDMA's Reliable Connection (RC), which provides reliable delivery with hardware-assisted timeouts and retransmission.

\subsection{Control and Data Planes}
\vspace{-2mm}
\label{sec:split-cp-dp}

AWs and EWs in \name exchange both data (token embeddings) and small control messages required for fault management. 
To avoid interference, \name 
allocates two Queue Pairs (QPs) per AW-EW pair: 
(1) a {\it control-plane QP} for liveness probes and self-healing metadata for rerouting and replay (\S\ref{sec:aw-side-self-healing}--\S\ref{sec:ew-side-self-healing}), and %
(2) a {\it data-plane QP} dedicated to bulk token embedding transfers. %
REFE uses the data-plane QP with GPUDirect RDMA to stream tensors directly into GPU memory, bypassing the CPU,
for higher performance.

\subsection{Reconfigurable Forwarding Engine}
\vspace{-2mm}
\label{sec:refe-design}

The Reconfigurable Forwarding Engine (REFE) is an AW-side runtime that coordinates point-to-point communication with EWs and orchestrates routing  during inference.
It exposes a simple API, \textit{expert\_io(expert\_id, layer\_id, token\_embeddings)}, to abstract the underlying RDMA control and data planes. %
The compute engine invokes the API after completing its attention computation.
Internally, REFE runs a non-blocking, event-driven execution loop
that takes output tokens, %
consults the ERT, and dispatches metadata and token embeddings to the selected EWs.

Beyond request dispatching, REFE manages the reception of expert outputs from EWs and performs AW-side liveness probing. Missing or delayed EW responses are detected through these probes, triggering REFE's self-healing logic (\S\ref{sec:aw-side-self-healing}), which transparently replays requests to healthy EWs.

\name's routing %
operates over point-to-point RDMA connections, 
giving each AW full flexibility to route individual requests to any EW without the reconfiguration of a collective communicator.
This communication pattern resembles the M2N communication used in prior MoE systems~\cite{megascale-infer, deepserve, eaas,step-3}.

\textnb{Expert Routing Table (ERT):}
\name decouples %
expert identity from expert location through the Expert Routing Table (ERT).
The ERT maps each expert to one or more candidate EWs---potentially including shadow experts (\S\ref{sec:shadow-experts-design})---allowing immediate rerouting when an EW fails or when additional EWs are provisioned with new expert replicas.

This indirection is the foundation of \name's self-healing: routing adaptation becomes a localized remapping operation rather than a system-wide recovery.
Each AW maintains its own ERT, %
updated by the orchestrator as the cluster evolves, ensuring that dynamic routing, fault isolation, and reconfiguration all occur with minimal disruption to ongoing inference.

\section{Worker Failure Management}
\vspace{-2mm}
\label{sec:worker-failure-mgt}

\textnb{Lightweight Failure Detection.} 
We build a hybrid liveness detection mechanism in \name:
Tokens exchanged between AWs and EWs  over their data-plane
QPs serve as \emph{implicit} heartbeat signals.
If a data-plane connection remains silent for longer than a configured  interval, the worker treats 
this as a potential indication of a failure and issues an \emph{explicit} probe over the control-plane QP to confirm a peer's liveness.
This design avoids unnecessary probing under normal load but provides fast detection when failures occur 
(see additional implementation details in Appendix~\ref{sec:detection-mechanism}).

\subsection{How Does AW Tolerate EW Failures?}
\vspace{-2mm}
\label{sec:aw-side-self-healing}

After an AW dispatches token embeddings to a selected EW, it waits for a response only for a bounded period.
If the EW fails to respond within the timeout, 
REFE probes the loss of liveness and on detecting a failure, immediately reroutes the request to an alternate EW hosting the same expert (either a healthy primary or a shadow expert, \S\ref{sec:shadow-experts-design}).

Because expert computation is stateless and deterministic, replaying the same metadata and token embeddings produces identical results. 
This design allows REFE to mask EW failures immediately---without waiting for the orchestrator to trigger global recovery---while preserving uninterrupted inference execution. %
Replayed requests are prioritized at the destination EW to prevent recovering AWs from becoming stragglers.
With this \emph{AW-side self-healing}, EW failures no longer manifest as global synchronization barriers, as shown in Fig.~\ref{fig:failure-case-studies}(c): only AWs that were issuing requests to the failed EW perform local rerouting and replay, while other workers continue to make forward progress.

\subsection{How Does EW Tolerate AW Failures?}
\vspace{-2mm}
\label{sec:ew-side-self-healing}
In \name, EWs tolerate AW failures by starting expert computation once
a \emph{sufficient subset} of AWs has delivered their tokens, proceeding
with partial inputs instead of waiting for all AWs to respond.
Concretely, for each expert and layer, an EW buffers incoming
tokens and starts the expert computation when either 
(i) it has received inputs from all AWs that are currently deemed healthy, or
(ii) the buffered batch reaches a configured minimum size that preserves GPU efficiency.%
\footnote{Expert kernels reach near-optimal GPU efficiency at moderate batch sizes~\cite{megascale-infer}.
As we evaluated in Appendix~\ref{sec:expert-batch-size}, once the expert batch
exceeds a modest threshold, GPUs sustain high efficiency. 
Executing with slightly fewer tokens is thus an acceptable trade-off:
it preserves GPU efficiency while avoiding the long tail latency induced by waiting for a failed AW.}
For any AW that has not contributed inputs within a short probing window,
the EW issues liveness probes; if the AW is still unresponsive, it is treated as having failed for this layer, and its slots are simply omitted from the current batch.
This \emph{EW-side self-healing} design removes the implicit global barrier of prior MoE systems, where every EW must wait for inputs from \emph{all} AWs before starting expert computation.

\subsection{Fast Expert Recovery via Shadow Experts}
\vspace{-2mm}
\label{sec:shadow-experts-design}
EW failures reduce the system's expert capacity: when an EW crashes, all experts it hosts become unavailable until a replacement EW is provisioned and expert weights are reloaded, which can take hundreds of milliseconds to seconds~\cite{serverlessLLM}.

\name introduces shadow experts that are pre-loaded but are normally \emph{inactive} replicas of experts residing in EWs' GPU memory.
A shadow expert contains the same model weights and computation kernel as its primary, and can be activated immediately when the primary becomes unavailable.
This allows \name to restore lost expert capacity almost instantaneously while the orchestrator provisions replacement EWs in parallel.
We keep shadow experts inactive in the common case;
keeping multiple experts active on the same GPU can cause kernel-level interference~\cite{nvidia-mps}, 
increasing the expert's execution latency (refer to Appendix~\ref{sec:why-not-active-shadow-experts} for additional analysis on the latency of having the shadow experts remain idle vs. executing in parallel).

\smallskip
\textnb{GPU memory cost of shadow experts.}
Unlike AWs, EWs do not store the KV cache, so their footprint prior to activation is small.
The main cost of shadow experts is GPU memory, which could be modest in practice.
For example, a single DeepSeek-R1 expert occupies roughly 2.5~GB~\cite{deepseekr1} for its weights, a relatively small proportion of the 40 to 80~GB capacity of GPUs such as an NVIDIA A100.
Even with multiple active and shadow experts co-located, typical MoE configurations comfortably fit within device memory.

\subsection{New Worker Provisioning}
\vspace{-2mm}
\label{sec:new-worker-provision}

Introducing a new worker into the cluster includes:
(i) connecting it to the existing AW-EW datapath,
and (ii) synchronizing the new worker's frontier (initialized at \emph{layer~1}) with existing workers' without interrupting their progress.
A naive solution would let the orchestrator enforce a global frontier update, which is, however, hard to coordinate at sub-millisecond per-layer execution time (0.85~ms in Table~\ref{tab:empirical-params}).

\begin{figure}[b]
\vspace{-5mm}
\centering
    \includegraphics[width=\linewidth]{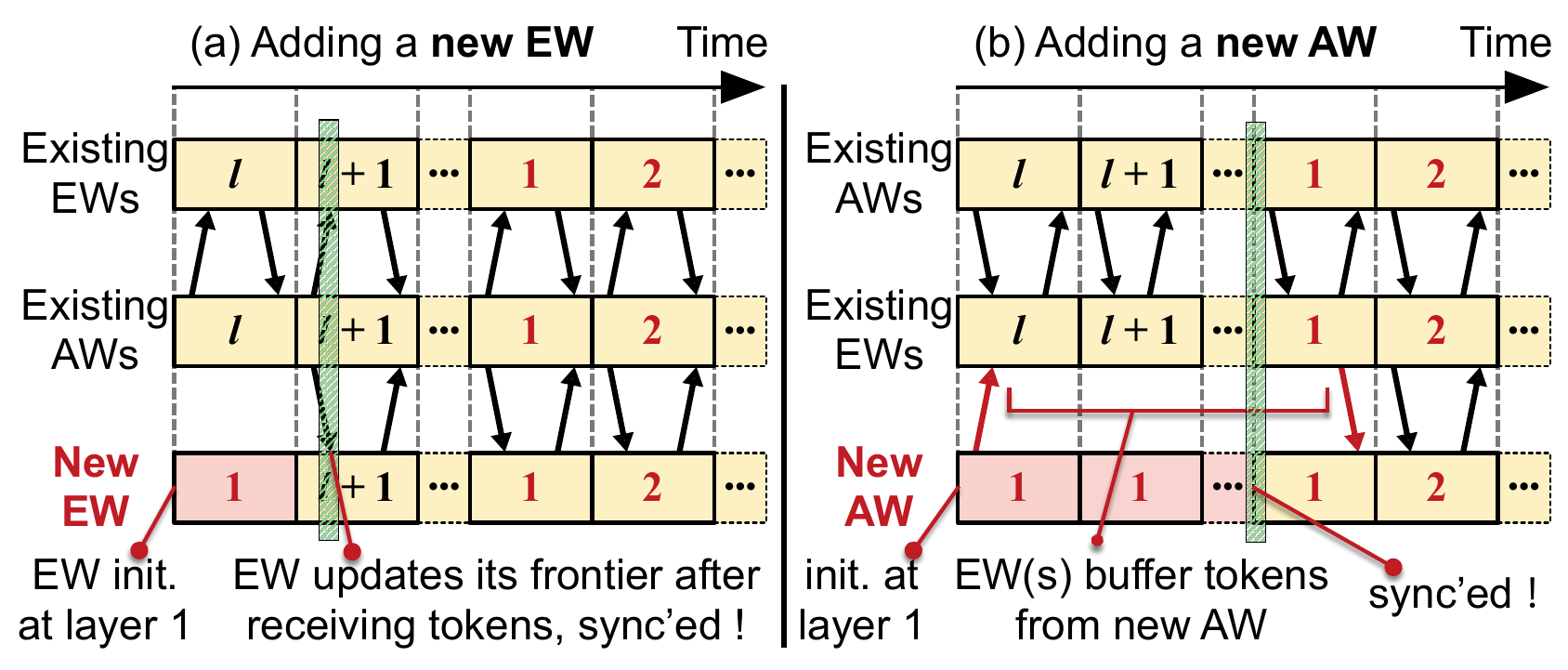}
\vspace{-8mm}
    \caption{New (a) EW and (b) AW provisioning. Numbers inside the boxes denote the layer index.}
    \label{fig:new-worker-provisioning}
\end{figure}

\smallskip
\textnb{Adding a new EW.}
When a new EW joins, the orchestrator updates all AWs' ERTs. Each AW then sets up the datapath to the new EW.
After connecting to all existing AWs, the new EW broadcasts a ready signal.
From this point on, AWs may start routing tokens to the new EW whenever its experts are selected.
Because existing AWs are already synchronized layer-wise, the first token the EW sees necessarily corresponds to the global frontier.
As such, the EW updates its local frontier according to the layer index carried in the first token's metadata (see Fig.~\ref{fig:aw-ew-datapath}), and then advances in lock-step with subsequent requests, as shown in Fig.~\ref{fig:new-worker-provisioning}(a).

\smallskip
\textnb{Adding a new AW.}
The principle of adding a new AW is similar.
The new AW boots and registers with the orchestrator to update its ERT.
It then sets up the datapath to all EWs and it can immediately start serving \emph{new} requests.
However, when EWs first receive tokens from this AW, their own frontiers may be at some layer $\ell \neq 1$ for the current token.
To preserve efficient layer-wise batching, EWs treat tokens from the new AW as follows:
as shown in Fig.~\ref{fig:new-worker-provisioning}(b), they buffer these ``early'' tokens until they themselves wrap back to layer~1 for the corresponding experts, and then batch the buffered tokens together with layer-1 tokens from other AWs.
From that point onward, the new AW is naturally synchronized at layer boundaries with the rest of the cluster.

This design (1) avoids global stalls: healthy workers never stop solely to accommodate a joining worker, unlike the coarse-grained recovery in Fig.~\ref{fig:failure-case-studies}; (2) preserves EW batching efficiency by letting both new EWs and new AWs join at layer-aligned points (next occurrence of layer~1).

\section{KV Cache State Management}
\vspace{-2mm}
\label{sec:kvcache}

\subsection{KV Cache Checkpointing}
\vspace{-2mm}
\label{sec:async_checkpoint}

\smallskip
\textnb{Initialization.}
When an AW starts, its compute engine allocates a contiguous KV cache region in GPU memory.
REFE registers this region for RDMA, enabling one-sided remote writes into the checkpoint store.
The AW then establishes an RDMA connection to the checkpoint store and sends the  
layout of its requested KV cache region (base address on the AW side and size).
The checkpoint store allocates and registers a dedicated memory bucket for this AW and returns the base address on the checkpoint store side.
Because both sides allocate fixed contiguous buffers, the AW can compute the remote write offset for 
the new KV cache, enabling direct one-sided updates without any receiver-side CPU involvement.

\smallskip
\textnb{Asynchronous incremental updates.}
For every layer, the AW incrementally updates one KV segment for each token. 
When the KV cache segment is updated,
the compute engine notifies REFE through a non-blocking \texttt{async\_update}() call, passing the address range of the new segment.
REFE then asynchronously issues a one-sided RDMA write that transfers this segment into the AW's bucket in the checkpoint store.

One-sided RDMA writes make checkpointing scalable but do not guarantee that segments arrive in order at the checkpoint store;
Later segments may arrive before earlier ones, making the checkpoint unusable for recovery.
To enforce ordering, \name adopts a standard ``async log + commit record'' design, similar to RDMA-backed write-ahead logging in prior systems~\cite{farm, herd, drtm}.
Each update is tagged with a \textit{sequence number} to preserve ordering, implemented as
a monotonically increasing RDMA work request ID.

\begin{figure}[t]
\centering
    \includegraphics[width=.9\columnwidth]{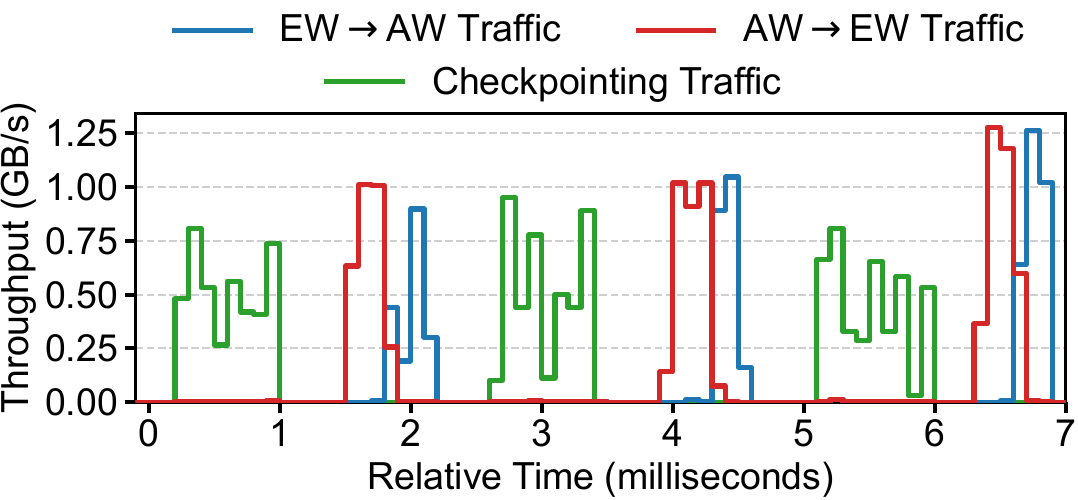}
\vspace{-4mm}
    \caption{Traffic pattern with incremental checkpointing.
    }
\vspace{-5mm}
    \label{res:checkpoint_traffic_pattern}
\end{figure}

\smallskip
\textnb{Opportunistic interleaving with AW-EW traffic.}
The checkpointing mechanism must run without interfering with the AW-EW traffic that drives the inference progress.
In practice, we find that AW-EW communication is
highly \emph{bursty}, as shown in Fig.~\ref{res:checkpoint_traffic_pattern}.
This is measured using the Mixtral-8$\times$7B model. Refer to \S\ref{sec:cr-test} for testbed setup.
Here we show a trace for a request arrival rate of 10~RPS, but we observe a similar pattern at other rates:
the link is heavily used when AWs scatter/gather expert inputs and outputs, but remains largely idle while AWs execute attention computation for the same layer.
These recurring idle intervals provide natural windows for incremental KV cache updates, that don't contend with AW-EW traffic. 
\name schedules incremental KV cache checkpoints just within these gaps.

\subsection{Request-Level KV Cache Restoration}
\vspace{-2mm}
\label{sec:kv-restoration}

\name recovers only the affected requests through the per-request KV cache restoration.
Upon detecting an AW failure, the orchestrator identifies all active requests on the failed AW by noting the latest committed token stored in the checkpoint store.
The orchestrator redistributes these requests to alternate AWs, typically in a round-robin manner to balance the load.

For each reassigned request, recovery proceeds as follows. The checkpoint store sends the alternate AW the request's committed token ID and the size of KV state to restore. The AW allocates a fresh per-request region in its KV cache and returns the offset to the checkpoint store.
Using GPUDirect-based one-sided RDMA writes, the checkpoint store then injects the KV cache segments directly into the AW's GPU memory,
followed by an 
HTTP request to confirm completion.

After the KV cache is reconstructed, the alternate AW resumes decoding from the committed token as if the request had been executed locally. 
Restoration is fully parallelized with the ongoing inference, because each request maintains its own isolated region. 
As a result, \name recovers AW failures without global rollback and without disrupting the progress of unrelated requests.

\begin{figure*}[htbp]
\centering
    \includegraphics[width=\linewidth]{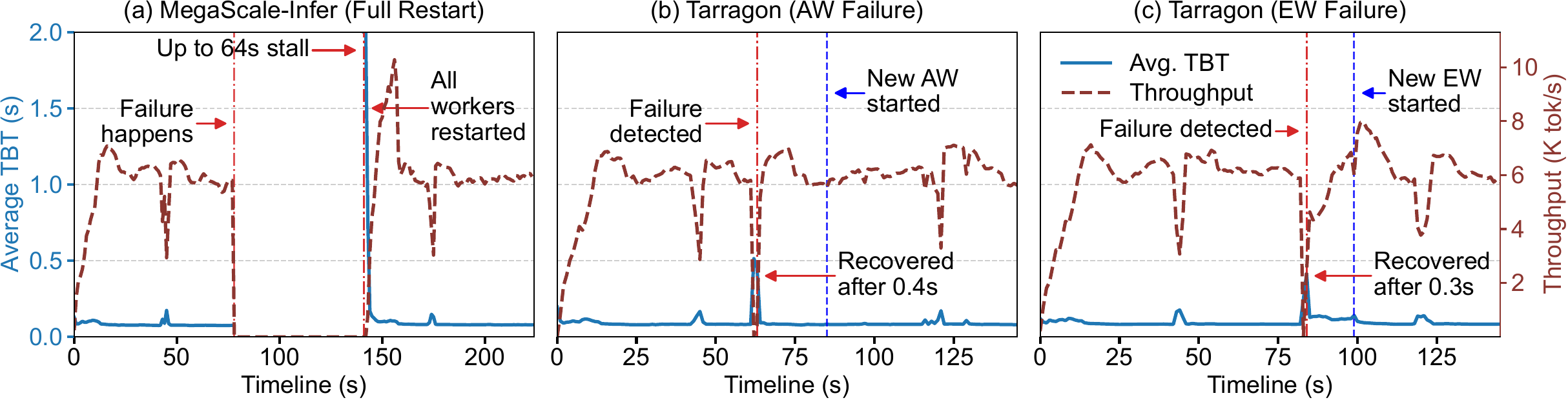}
\vspace{-8mm}
    \caption{End-to-end failover behavior in terms of time-between-tokens (TBT) and output tokens per second under a single worker failure. Note that (a) uses a longer time range than (b) and (c) because the MegaScale-style baseline experiences a much longer stall; thus, we run it longer to ensure performance has fully recovered after failover.}
\vspace{-6mm}
    \label{res:e2e-failover}
\end{figure*}

\section{Evaluation}
\vspace{-2mm}
\label{sec:eval}

Our evaluation is based on measurements of our implementation of \name running in Google Cloud.\\
\textnb{Implementation.}
We implement \name (about 16K lines of C++ code and 2K lines of Python).
For the AW, we use vLLM~\cite{vllm}  as the compute engine for both prefill and decoding, with the REFE implemented as a C++ extension with a Python shim.
EW is written from scratch in C++ using libtorch (PyTorch's C++ API) for expert computation and libibverbs for RDMA.
The orchestrator is a C++ control plane service exposing HTTP endpoints for configuration and failure monitoring of workers.
The checkpoint store is implemented as a separate C++ service using libibverbs.
We will open source soon.

\noindent
We aim to answer the following questions:
\begin{itemize}[leftmargin=10pt,nosep]
    \item How much can \name reduce user-visible stalls under worker failures compared to coarse-grained restarts? (\S\ref{sec:e2e-failover})
    \item What is the steady-state cost of adding failure resiliency, in terms of throughput and token-level latency, when no failures occur? (\S\ref{sec:cost-of-failure-resiliency})
    \item How much overhead do \name's individual resiliency components introduce by themselves? (\S\ref{sec:eval-ablation})
    \item How effective and lightweight are \name's KV cache checkpointing and restoration mechanisms? (\S\ref{sec:cr-test})
\end{itemize}
We compare \name against two state-of-the-art MoE serving frameworks, (monolithic) vLLM~\cite{vllm} and (decoupled) MegaScale-Infer~\cite{megascale-infer}. Both don't have fine-grained resiliency and have to rely on coarse-grained restarts after failures.

\subsection{Experimental Setup}
\vspace{-2mm}
\label{sec:expt-setup}

\textnb{Testbed:}
Unless otherwise noted, our experiments run on three Google Cloud (GCP) A3 Ultra nodes.
Each node has 224~vCPUs, 3~TB RAM, eight H200 GPUs (141~GB memory), and eight 400~Gbps ConnectX-7 RDMA NICs with GPUDirect RDMA and intra-node NVLink (3.6~Tbps) enabled.
All experiments run Ubuntu~22.04 with Linux~5.15, CUDA~12.8 (driver~580), and PyTorch~2.6.0.

\smallskip
\textnb{Model and workloads.}
We evaluate \name on Mixtral-8$\times$7B~\cite{jiang2024mixtralexperts}, a 32-layer MoE transformer with 8 experts per MoE layer and top-2 experts selected.
We use two prompt-completion workloads: ShareGPT~\cite{sharegpt}, with naturally varying input prompt lengths that examine both prefill and decode with realistic request heterogeneity; and a synthetic workload with randomly generated fixed-length prompts (10 input tokens, 128 generated tokens) to emphasize the decoding phase (called ``Random'').
Request arrivals follow a Poisson process with varying rates to emulate different load levels.

\smallskip
\textnb{Configuration of Baselines:}
For MegaScale-Infer, we follow its decoupled design and place 8 AWs on one A3 Ultra node and 8 EWs on the other node, so that all AW-EW traffic traverses inter-node RDMA links.
To match MegaScale-Infer's setting, \name uses the same 16 GPUs (8 AWs + 8 EWs on the first two nodes) and an additional node only for the checkpoint store, so it does not benefit from extra GPU capacity.
For monolithic vLLM, we adopt the two standard configurations that best exploit its design: vLLM-TP with tensor parallelism degree 16 and vLLM-PP with a 16-stage pipeline.
Both vLLM baselines are deployed on two A3 Ultra nodes with a total of 16 GPUs, leveraging intra-node NVLink for fast GPU-GPU communication (the recommended practice for TP/PP-style multi-GPU inference).
All systems share the same backend kernels and batching policy for a fair comparison, and disable any optimizations not common to all.
\name does failure probing every 10\,ms; the baselines do not have failure detection.

\begin{figure*}[htbp]
\centering
    \includegraphics[width=\linewidth]{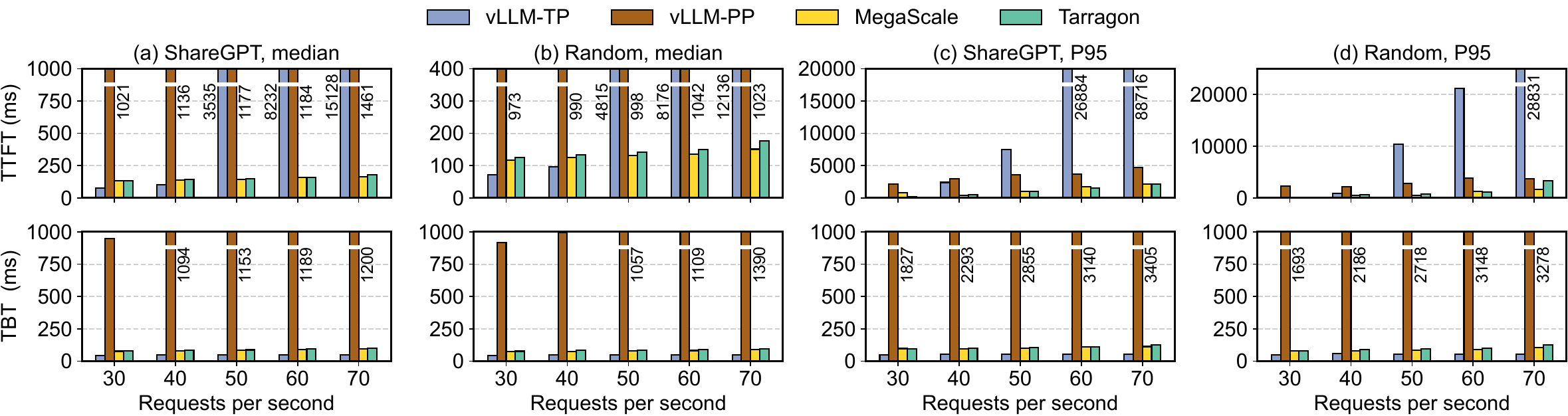}
\vspace{-8mm}
    \caption{Cost of failure resiliency on token-level latency. Top: TTFT. Bottom: TBT. We report medians (a-b) and P95 (c-d) across ShareGPT and Random workloads as load increases (30-70 RPS).}
\vspace{-6mm}
    \label{fig:ttft-tbt-no-failure}
\end{figure*}

\subsection{End-to-End Failover Behavior}
\vspace{-2mm}
\label{sec:e2e-failover}

\textnb{Setup:}
We first focus on failover behavior in the decoding phase, during which the model emits tokens under tight latency constraints.
We use the ``Random'' workload, which stresses the decoding phase, and set  the request arrival rate to 50 RPS  to keep the system under moderate load, so that observed stalls mainly reflect failover behavior rather than overload.
We compare \name against MegaScale-Infer.
Our primary metrics are the \emph{time between tokens} (TBT) and inference throughput (output tokens per second).
Increased TBT during failures will be perceived by users as the disruption in the stream of %
responses.

For each run, we start a long-running interactive request stream  
and let the pipeline reach the steady-state decoding phase.
Around 60–80\,s after the first request is issued, we inject a fail-stop worker failure by sending \texttt{SIGINT} to one worker process. 
For \name, we evaluate two failure scenarios separately:
(i) failure of a single AW, and (ii) failure of a single EW.
In both cases, \name's self-healing mechanisms will provision a new worker 
in the background.

\textnb{Results:}
Fig.~\ref{res:e2e-failover}(a) shows the TBT and throughput timeline for the MegaScale-Infer. When the failure is injected at 78\,s, 
the throughput immediately drops to \textit{zero}. The system kills and restarts all workers, reloads model weights, and reruns both prefill and decoding before it can emit the next token.
The stall lasts for roughly $64$\,s, consistent with our cost model in \S\ref{sec:cost-analysis}, and is visible to the user as a frozen response stream.

Figs.~\ref{res:e2e-failover}(b) and~(c) show the corresponding behavior under \name. When an AW fails, \name's self-healing reroutes the affected request to healthy AWs and replays only the minimal state needed at the current frontier. The resulting stall (and period during which throughput drops) is only about 0.4\,s, a 160\X reduction compared to the baseline. When an EW fails, \name masks the failure by replaying expert computation to healthy EWs and shadow experts, while a replacement EW is provisioned in the background. While waiting for the replacement EW, the reduced capacity results in a slightly elevated TBT. But, the actual stall in the token stream is substantially reduced, to just about 0.3\,s, or 213\X shorter than the baseline. In both cases, token generation resumes more quickly compared to the minute-long pause.
After the new EW's initialization, it  joins the cluster. The TBT returns to its pre-failure level.
This behavior is precisely what \name is designed to achieve: \textit{self-healing hides the long worker initialization latency from users, while background provisioning fairly quickly restores the original capacity}.

\begin{figure}[b]
\vspace{-4mm}
\centering
    \includegraphics[width=\columnwidth]{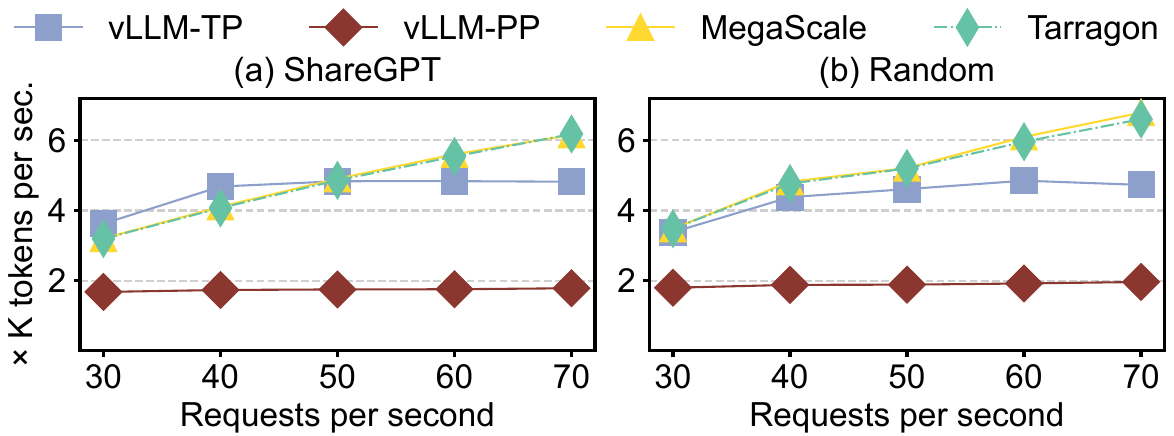}
\vspace{-8mm}
    \caption{Cost of failure resiliency on output tokens per second (higher is better) for (a) ShareGPT and (b) Random.}
    \label{fig:output-tok-no-failure}
\end{figure}

\subsection{Is There a Cost to Failure Resiliency?}
\vspace{-2mm}
\label{sec:cost-of-failure-resiliency}

\textnb{Setup:} We now compare \name against three non-resilient baselines:
vLLM-TP, vLLM-PP, and MegaScale-Infer, under non-failure conditions. 
We vary the load from 30 to 70 RPS and report (i) TTFT (time-to-first-token) and TBT (median and P95), (ii) output-token throughput for both ShareGPT and Random workloads.

\smallskip
\textnb{TTFT (Fig.~\ref{fig:ttft-tbt-no-failure}, top row)}. 
For prefill,  \name closely tracks MegaScale across loads on both workloads, indicating that \name's resiliency adds negligible startup latency.
But, the two vLLM baselines show different behaviors.
At low to moderate load (30-40~RPS), vLLM-TP achieves slightly lower TTFT than the decoupled systems.
This is helped by its use of high-bandwidth NVLink for intra-node communication, so prefill completes quickly when the cluster is not saturated.
However, as load increases beyond 40~RPS, vLLM-TP's TTFT grows very sharply, reaching  multi-second delays.
vLLM-PP exhibits consistently worse TTFT than both MegaScale and \name at all loads.

\smallskip
\textnb{TBT (Fig.~\ref{fig:ttft-tbt-no-failure}, bottom row)}.
For decoding, \name and MegaScale again remain close. 
The differences among vLLM baselines are mainly driven by how well their parallelism strategies fit autoregressive decoding.
vLLM-PP shows substantially larger TBT across all loads.
vLLM-TP generally achieves slightly better TBT than the decoupled systems.
By splitting each transformer layer across %
GPUs and using NVLink-backed collectives to gather partial results, vLLM-TP can keep its GPUs busy and hide much of the intra-node communication cost.
In contrast, MegaScale and \name must perform AW-EW scatter/gather over inter-node RDMA links; this additional network hop introduces a small but visible latency penalty per token.

\smallskip
\textnb{Output-Token Throughput (Fig.~\ref{fig:output-tok-no-failure}).}
\name essentially matches MegaScale's throughput (with a deviation within 2.8\%).
vLLM-PP and vLLM-TP deliver consistently lower throughput: pipeline parallelism leaves some GPUs underutilized due to pipeline bubbles and imbalance, while tensor parallelism pays per-layer collective overhead across the GPUs, which limits the effective throughput at high load.

\smallskip
\textnb{Takeaway.}
These results demonstrate that \name preserves the performance benefits of decoupled deployment, matching MegaScale-Infer in latency and throughput across various loads and workloads. It outperforms monolithic vLLM-TP/PP in most settings—while adding strong failure resiliency with little or no additional cost in the no-failure common case.

We also conduct an ablation study to quantify the steady-state overhead of incremental KV cache checkpointing, lightweight failure detection, and ERT-based expert remapping, on \name's end-to-end throughput without injecting failures. The results showed that across all request rates and workloads, the throughput of all alternatives are nearly indistinguishable: the maximum difference is less than 3\% (details are in Appendix~\ref{sec:eval-ablation}).

\subsection{KV cache Checkpointing and Restoration}
\vspace{-2mm}
\label{sec:cr-test}

We perform this experiment on three GCP A3 Ultra nodes (\S\ref{sec:expt-setup}), where AWs, EWs, and checkpoint stores run on separate nodes with RDMA NICs interconnecting them.

\smallskip
\textnb{Overhead of different checkpointing schemes.}
Our goal here is to understand whether different checkpointing schemes interfere with the inference, in particular whether \name's can preserve inference throughput while providing fine-grained checkpointing.
We consider two baselines: (1) \textit{No checkpointing}, which serves as the upper-bound baseline; (2) \textit{Pause-Checkpoint-Resume}, 
which periodically stalls inference to take a global snapshot of \emph{all} KV cache pages (the training-style approach), and then resumes decoding.

For \emph{Pause-Checkpoint-Resume}, we vary the checkpoint interval by the number of generated tokens: after every $X$ decoded tokens, the system pauses, checkpoints the whole KV cache, and then resumes. 
\name does not use such periodic intervals. 
In \name, once a token's KV segment has been updated, the AW immediately checkpoints it, during which AW-EW traffic is naturally idle (Fig.~\ref{res:checkpoint_traffic_pattern}).

We report end-to-end inference throughput in output tokens per second.
Without checkpointing, the system achieves 1148~tokens/s.
With \name's asynchronous incremental checkpointing, throughput is 1147~tokens/s, essentially identical to \emph{no checkpointing}.
This confirms that opportunistic interleaving is effective: KV cache updates occur during link idle periods and do not measurably interfere with normal AW-EW communication.

In contrast, \emph{Pause-Checkpoint-Resume} incurs significant overhead even at relatively coarse intervals. With a checkpointing interval of once every 8 tokens, throughput drops by 2.15$\times$ compared to both \emph{no checkpointing} and \name. The degradation stems from repeated global stalls: each checkpointing pauses the entire pipeline, blocking new token generation while KV-state is flushed.
Achieving token-level checkpointing with \emph{Pause-Checkpoint-Resume} would require even more frequent stalls and thus be prohibitive for interactive inference.

\begin{figure}[t]
\centering
    \includegraphics[width=\columnwidth]{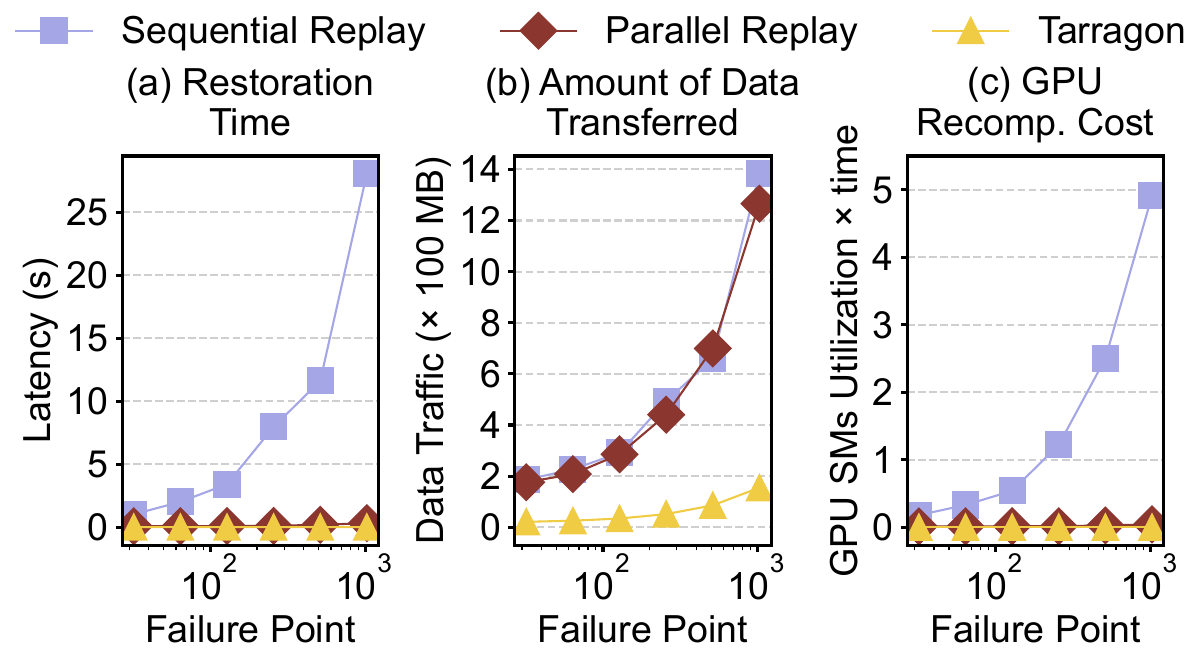}
\vspace{-8mm}
    \caption{Impact of different restoration strategies at varying failure points.}
\vspace{-5mm}
    \label{res:restoration-eval}
\end{figure}

\smallskip
\textnb{AW restoration.}
We evaluate how \name's per-request KV restoration impacts AW-side self-healing during decoding-time failures. 
We focus on a single AW failure during the decoding phase (since it dominates user-perceived impact) and compare against two replay-based baselines:
(1) \textit{Sequential replay}: an alternate AW rebuilds the lost KV cache by rerunning prefill and then \emph{sequentially} decodes all tokens from the beginning up to the failure point, without using any checkpoints.
(2) \textit{Parallel replay}: an alternate AW performs the prefill over the original prompt plus all tokens generated up to the failure point, reconstructing the lost KV cache in \emph{parallel} rather than token by token.

We vary the \emph{failure point}, which is defined as the index of the token being decoded when the failure occurs.
A larger failure point corresponds to a larger number of decoded tokens and a larger KV cache to recover.
For each strategy, we measure: (1) the total restoration time, (2) the amount of data transferred (\ie AW-EW traffic for \textit{sequential replay} and \textit{parallel replay}, and AW--checkpoint-store traffic for \name), and (3) the GPU recomputation cost (GPU-time) incurred by the alternate AW.

Fig.~\ref{res:restoration-eval}(a--c) summarizes the results.
As the failure point increases, \textit{sequential replay} has a steep increase in all three metrics: restoration time and GPU-time both grow roughly linearly with the increase of failure point, since the alternate AW must rerun attention for every layer and token, resulting in additional AW-EW traffic.
\textit{Parallel replay} 
still incurs the same amount of AW-EW traffic as \textit{sequential replay}, growing rapidly with the failure point. \textit{Parallel replay} incurs smaller restoration delays than \textit{sequential replay}, but is still roughly an order of magnitude higher than \name (10\X).

It is important to see that \name's per-request restoration remains nearly constant and efficient across all tested failure points.
GPU recomputation cost and total restoration time are both negligible, as no prefill/decoding work is replayed.
The amount of data transferred in \name is roughly $1/8$ of that in \textit{sequential replay} and \textit{parallel replay}.
Restoring the KV cache for a single request is far cheaper than regenerating it (up to 1800\X latency reduction compared to \textit{sequential replay}).
As a result, AW-side self-healing under \name can recover from failures quickly and in an isolated manner, without flooding the network or burning scarce GPU cycles.
This complements the end-to-end failover results in \S\ref{sec:e2e-failover}, showing that \name not only hides failures at the token-processing level but also bounds the system-wide recovery cost.

\section{Related Work}
\vspace{-2mm}
\label{sec:related-work}

\textnb{Optimizations on MoE serving.}
Recent work on MoE serving has primarily focused on efficiency---reducing resource usage, improving GPU utilization, and lowering communication overhead~\cite{megascale-infer,expert-buffering-2024,speculative-moe-2025,deepseekv3-tr,step-3}. These systems, however, retain static expert placement, fixed communication groups, and tightly synchronized AW-EW execution, causing a single worker failure to trigger coarse-grained restarts with no worker-level failover. \name complements these efficiency-oriented designs by introducing a reconfigurable datapath and bidirectional self-healing that confine and recover AW and EW failures at worker granularity.

\smallskip
\textnb{Fault tolerance in LLM serving.}
Existing resilient serving approaches primarily target predictable events. For instance,
SpotServe~\cite{spotserve} adapts parallelism during preemption windows on cloud spot instances but does not handle sudden failures nor exploit the structure of decoupled attention-expert deployments.
For KV cache durability and scalability, systems such as MoonCake~\cite{Mooncake} build a distributed KV cache store. \name is orthogonal to these efforts: it provides fine-grained failover for both AWs and EWs and can optionally integrate such a store to further strengthen KV cache recovery. %

\smallskip
\textnb{Fault tolerance in LLM training.}
Resilient training systems address failures through checkpointing~\cite{varuna}, expert replica placement strategies~\cite{lazarus-duke}, redundant computations~\cite{bamboo}, and state reconstruction from healthy replicas~\cite{oobleck}.
These techniques are effective for long-running, globally synchronized training jobs, 
but operate in a fundamentally different regime from inference, which must recover within tight latency budgets and preserve KV cache state.

\section{Conclusion}
\vspace{-2mm}
Based on understanding the deficiencies of coarse-grained failover in today's MoE serving frameworks, we developed \name. \name confines failure domains to individual workers, maintains forward progress of inference pipelines under failures, and limits the amount of processing required for recovery.  
Our evaluation shows that \name cuts failure-induced stalls by up to 160–213\X compared to state-of-the-art MoE serving frameworks, while matching their throughput 
when no failures occur.
We believe these results demonstrate that strong failure resilience and high-performance MoE serving are not at odds, and that \name provides a practical solution for making large-scale LLM inference robust to the routine GPU and node failures seen in production clusters.

\bibliographystyle{plain}
\bibliography{reference}

\clearpage

\appendix

\setcounter{page}{1}

\section{Failure Propagation in Monolithic and Decoupled Deployments}\label{sec:case-studies-walk-through}

\subsection{Case study I: monolithic deployment mode}
Fig.~\ref{fig:failure-case-studies} (a) presents how a single worker's failure propagates in a monolithic deployment. 
When \emph{Worker 1} fails during decoding,
the collective communicator (CCL) is torn down after a timeout.
Importantly, the CCL treats the set of workers as a static communication group~\cite{Hoplite}: if one worker becomes unavailable, the entire communicator aborts.
Consequently, all workers are killed and restarted and the CCL is re-initialized.
Then all workers must replay all $L$ prefill layers, all decoding layers for the $(i-1)$ tokens that have already been produced, and finally the first $\ell$ layers for the current token have to be re-executed.
All previously accumulated KV caches and partial outputs---even on healthy workers---are discarded. This results in a long, user-visible inference stall and substantial recomputation overhead.

\subsection{Case study II: decoupled deployment mode}

\textit{AW failure:}
Fig.~\ref{fig:failure-case-studies} (b) shows the failure of a single AW. 
Upon failure, only the failed AW is restarted and reinitialized. Because the KV cache and frontier state for its requests are lost, the restarted AW must re-run the prefill and decode from the beginning (along with the corresponding EW computation) until it reaches the \textit{same frontier} (layer $\ell$ of the $i$-th token) as the rest of the pipeline. Meanwhile, EWs must wait at the layer-wise synchronization barrier, stalling the entire inference pipeline until the failed AW catches up.
Thus, despite preserving healthy workers, recovery still incurs long stalls and extensive recomputation, closely mirroring the monolithic case. The healthy workers remain running and merely wait at the synchronization barrier, without doing the recomputation.

\textit{EW failure:}
An EW failure behaves differently but still introduces an inference stall, as depicted in Fig.~\ref{fig:failure-case-studies}(c).
Since EWs are stateless, an EW failure does not result in discarding previously computed KV caches or decoding progress preserved on the AWs.
Thus, the failed EW needs to be restarted.
Once the replacement EW is ready,
it can recompute the expert output only at the current frontier $\ell$. 
However, the AW that routed tokens to the failed EW must wait at the current \emph{frontier} (layer $\ell$) until the replacement EW is back online. As a result, decoding still experiences a user-visible stall, despite minimal recomputation.

\section{Expert Batch Size}
\label{sec:expert-batch-size}

\begin{figure}[htbp]
\centering
    \includegraphics[width=\linewidth]{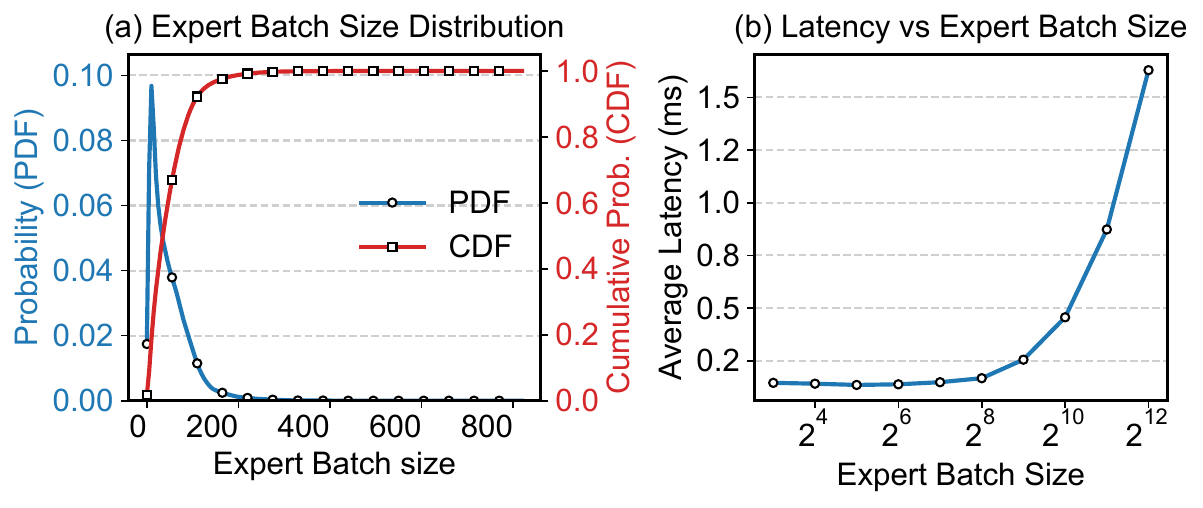}
    \vspace{-6mm}
    \caption{(a) Run total batch size of 821 and collect the expert batch size distribution of Qwen3-MoE~\cite{qwen3technicalreport} across different layers. (b) The single expert computation latency of Qwen3-MoE~\cite{qwen3technicalreport} on 8 x Nvidia A100 with different expert batch sizes.}
    \label{res:consolidated-moe-serving-microbench}
\end{figure}

Sparse expert activation also has a subtle but important side effect: it fragments tokens across many experts and leads to small per-expert batches.
In our measurements on Qwen3-MoE, even when the \textit{total} token batch size is 821, the vast majority of per-expert batches contain fewer than 200 tokens (Fig.~\ref{res:consolidated-moe-serving-microbench} (a)).
However, on an NVIDIA A100, expert kernels only reach their throughput ``knee point'' at batch sizes of roughly 256$\sim$512 tokens (\ie an exponential growth of latency in Fig.~\ref{res:consolidated-moe-serving-microbench} (b)), so most expert invocations run far below the GPU's efficient operating regime.
When attention layers and experts are co-located on the same GPUs, the memory-bound attention computation further constrains batch size, compounding underutilization (also observed in~\cite{megascale-infer}).

\section{Checkpointing Overhead Analysis}\label{sec:ckpt_overhead}
Checkpoint overhead is dominated by the size of each incremental KV cache segment, which is:
\[
C = 2 \times H_{\text{kv}} \times \frac{N_{\text{hidden\_size}}}{H_{\text{attn}}} \times S_{\text{elem}}
\]
where $N_{\text{hidden\_size}}$ is the hidden size and $S_{elem}$ is the number of bytes per tensor element. $H_{\text{kv}}$ and $H_{\text{attn}}$ are the numbers of KV heads and attention heads, respectively.

In memory-efficient attention mechanisms such as multi-query attention (MQA) and grouped-query attention (GQA)~\cite{multi-query-attention, gqa}, $H_{\text{kv}} \ll H_{\text{attn}}$, substantially reducing checkpointing traffic. By contrast, the per-token, per-layer communication volume between the AW and EW is:
\[
V = 2 \times \text{Top}_{k} \times N_{\text{hidden\_size}} \times S_{\text{elem}}
\]
which is significantly larger ($\text{Top}_{k}$ is the number of expert chosen per layer). 
For example, in Mixtral-8x7B~\cite{jiang2024mixtralexperts}, incremental KV cache traffic is only $\sim$12.5\% of expert traffic.

\section{Inactive Shadow Experts During Normal Execution}\label{sec:why-not-active-shadow-experts}

\begin{figure}[htbp]
\centering
    \includegraphics[width=\columnwidth]{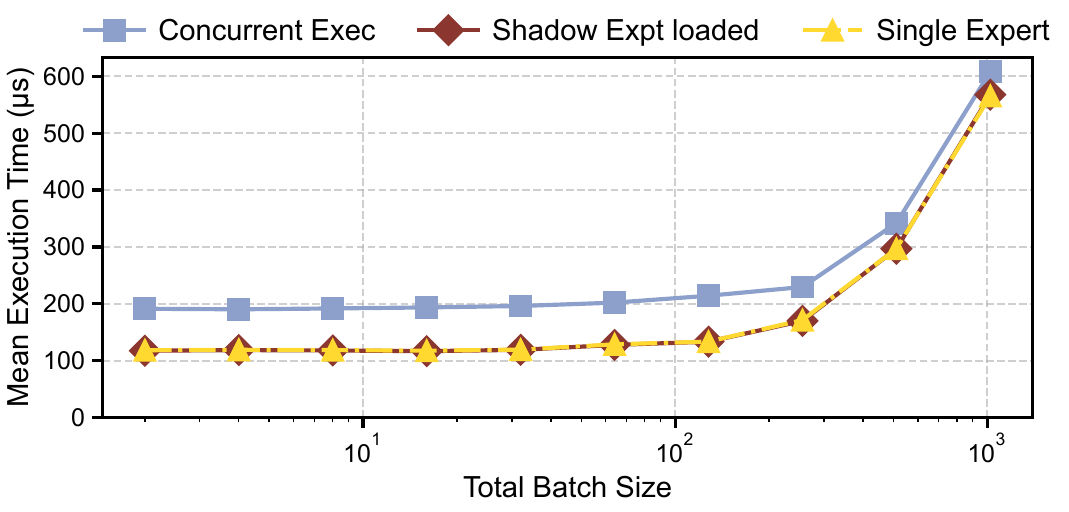}
\vspace{-8mm}
    \caption{Impact of shadow expert on execution latency}
\vspace{-4mm}
    \label{fig:shadow_expert_computation_efficiency}
\end{figure}

We find that executing two experts concurrently (\texttt{Concurrent Exec}) with two CUDA streams results in significant interference compared with activating only a single expert (\texttt{Single Expert}), using the same total batch size.
As shown in Fig.~\ref{fig:shadow_expert_computation_efficiency}, the computation latency of activating a single expert (\texttt{Single Expert}) is identical to the latency of loading an additional shadow expert, but not activating it (\texttt{Shadow Expt Loaded}).
Without failures, the loaded shadow expert consumes GPU memory but does not introduce any computational overhead for the activated expert.

\section{Implementation of Lightweight Failure Detection}
\label{sec:detection-mechanism}

We implement the explicit probes as ``zero-length'' RDMA writes, a no-op operation that incurs minimal overhead. %
To track the status of probes, each worker monitors its RDMA Completion Queue (CQ) to inspect the \texttt{ibv\_wc\_status} of the relevant QP (data-plane QP for the implicit probe and control-plane QP for the explicit probe).
If a probe
experiences a small number of consecutive timeouts (default value is 3, configured at QP initialization in \name), 
the RNIC marks the QP with \texttt{IBV\_WC\_RETRY\_EXC\_ERR}
(raises \texttt{IBV\_WC\_WR\_FLUSH\_ERR} for work requests in CQ), and flushes all pending work requests on that QP.
\name interprets these hardware-level signals as a fail-stop event on the
corresponding peer (either a worker or link failure) and immediately hands them to the recovery logic to trigger self-healing and worker replacement.

\section{Ablation Study}
\label{sec:eval-ablation}

We perform an ablation study on \name's main resiliency components to understand their individual contribution to overall performance, including: (i) incremental KV cache checkpointing (\S\ref{sec:async_checkpoint}) (ii) lightweight failure detection; (iii) the ERT used for dynamic expert remapping (\S\ref{sec:refe-design}).

We evaluate three variants of \name: (Alt-1) disables KV cache checkpointing/restoration; (Alt-2) additionally disables failure detection; (Alt-3) further disables the ERT.
We use the same model and workloads as in \S\ref{sec:cost-of-failure-resiliency} and vary the request arrival rate.
In this experiment, we \emph{do not inject failures} so that any performance differences are purely due to the steady-state overheads of these components. We report end-to-end inference throughput in terms of output tokens/sec.

\begin{figure}[htbp]
\centering
    \includegraphics[width=\columnwidth]{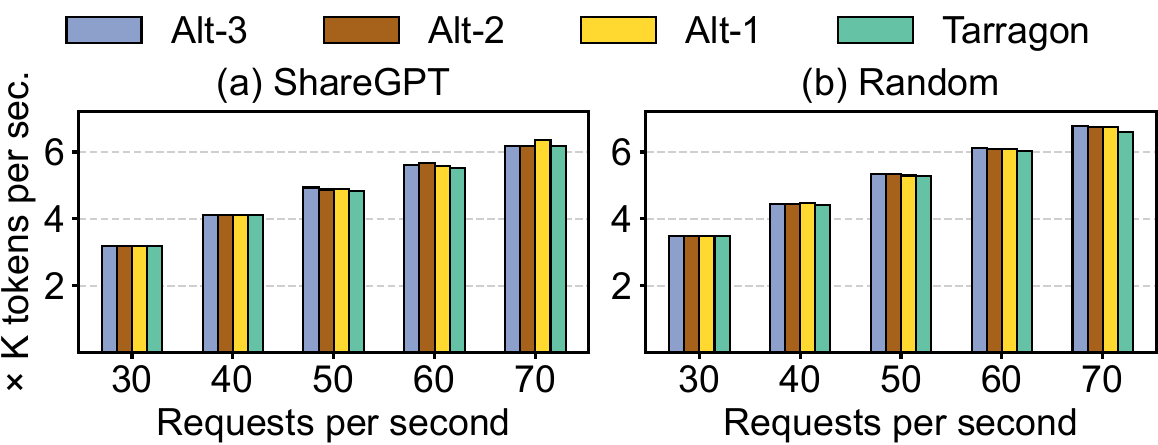}
\vspace{-8mm}
    \caption{Ablation study of \name's main resilience components. We evaluate output tokens per second for (a) ShareGPT and (b) Random.}
    \label{res:ablation_study}
\end{figure}

Fig.~\ref{res:ablation_study} shows that across all request rates and workloads, the throughput of all alternatives is nearly indistinguishable: the maximum deviation stays within 3\%.
In particular, \name with all components enabled matches the no-resiliency baseline (Alt-3, similar to MegaScale-Infer), for this no-failure case.

\end{document}